\documentclass[english,11pt,aps,prd,a4paper,preprintnumbers,floatfix,nofootinbib,showpacs,superscriptaddress, notitlepage]{revtex4-2} 
\usepackage[mode=buildnew]{standalone}
\usepackage[usenames,dvipsnames]{color}  
\usepackage{graphicx}

\usepackage{setspace}
\usepackage{caption}
\usepackage{subcaption}
\usepackage{amsmath}
\usepackage{amssymb}
\usepackage[colorlinks=true,citecolor=darkred,urlcolor=darkred, pdfborder={0 0 0}]{hyperref}
\usepackage[normalem]{ulem}
\usepackage{xcolor}
\usepackage{placeins}
\usepackage{mathrsfs}
\usepackage{verbatim} 
\usepackage{cancel} 
\usepackage{float} 

\usepackage{scalerel}
\usepackage{tikz-feynman}
\tikzfeynmanset{compat=1.1.0}
\usepackage{multirow}

\makeatletter
\def\p@subsection{}
\makeatother

\definecolor{darkred}{rgb}{0.6,0,0}

\definecolor{linkcolor}{rgb}{0,0,0.5}

\def\gsim{\raise0.3ex\hbox{$\;>$\kern-0.75em\raise-1.1ex\hbox{$\sim\;$}}}
\def\lsim{\raise0.3ex\hbox{$\;<$\kern-0.75em\raise-1.1ex\hbox{$\sim\;$}}}

\def\beqn#1{\begin{equation}\label{#1}}
\def\eeqn{\end{equation}}

\def\beqa#1{\begin{eqnarray}\label{#1}}
\def\eeqa{\end{eqnarray}}

\newcommand {\ignore}[1]{}

\def\Z4{$Z_4$}
\def\O5{$\mathcal{O}_5$ }

\def\321{$\mathrm{SU(3) \otimes SU(2) \otimes U(1)}$ }

\newcommand{\TBM}{\text{TBM}}

\begin{document}

\title{\color{Blue} {Leptonic CP phase correlations from modified TBM mixing and its modular $A_4$ realization in light of the latest JUNO results}}

\author{Sin Kyu Kang}
\email{skkang@seoultech.ac.kr}
\affiliation{School of Natural Science, Seoul National University of Science and Technology, Seoul 01811, Republic of Korea}
 \author{Ranjeet Kumar} \email{kumarranjeet.drk@gmail.com}
 \affiliation{Institute for Convergence of Basic Studies, Seoul National University of Science and Technology, Seoul 01811, Republic of Korea}

\begin{abstract}
\noindent
The improved determination of the solar mixing angle by the Jiangmen Underground Neutrino Observatory (JUNO), based on 207.2 live days of data, provides a more stringent test of lepton mixing sum rules and their Dirac CP phase predictions. We reassess four minimally modified tribimaximal mixing patterns (A--D) using JUNO and NuFIT 6.1. The correlated NuFIT profiles strongly disfavor the solar-reactor relation of case B but accommodate that of case A and the atmospheric-reactor relations of cases C and D in both mass orderings. A separate Gaussian comparison indicates increased tension for cases A and B with the updated JUNO result. 
For normal ordering, the updated JUNO measurement increases
the tension with case A without significantly improving
its CP phase prediction. In contrast, for cases C and D,
the minimum discrepancies with the oscillation data remain
unchanged, while the predicted CP phases shift by about
$2.3^\circ$ and their conditional intervals narrow by
approximately 25\%.
We demonstrate that these three mixing patterns can be
realized at their conditional minima within a previously
proposed modular $A_4$ seesaw model without modifying its
mass matrices.
The model also admits distinct neutrino mass spectra and
effective Majorana masses at fixed oscillation observables.
\end{abstract}

\maketitle

\section{Introduction}
\label{sec:intro}


The observation of neutrino oscillations established that neutrinos are massive~\cite{Kamiokande-II:1990wrs,Kamiokande-II:1992hns,Super-Kamiokande:1998kpq,Cleveland:1998nv,SNO:2002tuh}. This is in contrast to the minimal Standard Model (SM), in which neutrinos are massless, and thus provides compelling evidence for physics beyond the Standard Model (BSM).
In the three neutrino framework, lepton flavor mixing is described by the Pontecorvo-Maki-Nakagawa-Sakata (PMNS) matrix~\cite{Pontecorvo:1957cp,Maki:1962mu}, which is characterized by three mixing angles and one Dirac CP-violating phase relevant to neutrino oscillations. Despite substantial progress in determining the mixing angles, the origin of the observed lepton flavor structure, the Dirac CP phase, the octant of the atmospheric mixing angle, and the neutrino mass ordering remain central questions in neutrino physics.

The measurement of a nonzero reactor mixing angle, $\theta_{13}$, opened the possibility of probing leptonic CP violation in long-baseline neutrino oscillation experiments~\cite{DoubleChooz:2011ymz,DayaBay:2012fng} and provided an important test of previously proposed lepton mixing patterns~\cite{Barger:1998ta,Baltz:1998ey,Fritzsch:1998xs,Mohapatra:1998bp,Kang:1998gs,He:2006qd,Albright:2008rp,He:2011gb}. Although the exact tribimaximal (TBM) mixing pattern is incompatible with the measured nonzero value of $\theta_{13}$, it remains a useful leading order description when supplemented by suitable modifications~\cite{Kang:2014mka,Ahn:2022ufs}. Such modifications can accommodate the observed deviations from TBM mixing and lead to characteristic correlations among the physical mixing angles and the Dirac CP phase. These correlations, expressed as mixing sum rules, can be tested with precision neutrino oscillation measurements.

In light of the recent high-precision measurements of the solar oscillation parameters by the Jiangmen Underground Neutrino Observatory (JUNO)~\cite{JUNO:2026tap}, it is particularly timely to revisit these correlations, especially those involving the solar mixing angle, $\theta_{12}$.
In its first oscillation analysis, based on 59.1 live days of data, JUNO reported~\cite{JUNO:2025gmd}
\begin{equation}
    \sin^2\theta_{12}=0.3092\pm0.0087,
    \qquad
    \Delta m^2_{21}
    =(7.50\pm0.12)\times10^{-5}\ {\rm eV}^2,
    \label{eq:juno-first}
\end{equation}
for normal ordering (NO). The implications of these measurements for neutrino oscillations, lepton flavor symmetries, and neutrino mass models have been examined in a wide range of theoretical studies~\cite{Chao:2025sao,Li:2025hye,Huang:2025znh,Ge:2025cky,Xing:2025xte,Chen:2025afg,Petcov:2025aci,Goswami:2025wla,Ding:2025dqd,Borah:2025vtn,Nanda:2025fvw,Shang:2026qkh,Zhang:2025jnn,He:2025idv,Jiang:2025hvq,Ding:2025dzc,Dutta:2026dzh,Kumar:2026qee,Priya:2026tpk}.

JUNO has now reported an updated analysis based on 207.2 live days of data~\cite{JUNO:2026tap}, yielding\footnote{For recent phenomenological studies exploring the implications of the updated JUNO results, see Refs.~\cite{Alikhanov:2026bhi,Jin:2026ejo,Wang:2026zdw,Xing:2026dim,Machado:2026shm}.}
\begin{equation}
    \sin^2\theta_{12}=0.3036\pm0.0064,
    \qquad
    \Delta m^2_{21}
    =(7.388\pm0.078)\times10^{-5}\ {\rm eV}^2.
    \label{eq:juno-new}
\end{equation}
Compared with the first JUNO result, the updated measurement shifts the central value of $\sin^2\theta_{12}$ downward and reduces its uncertainty by approximately 26\%, providing a more stringent test of mixing patterns involving the solar mixing angle. The updated analysis also determines the atmospheric
mass-squared difference, $\Delta m^2_{31}$, with
approximately one-percent precision, using the Daya Bay
constraint to resolve the physical minimum under each mass ordering hypothesis. When combined with constraints from long-baseline accelerator experiments, the JUNO result favors NO over inverted ordering (IO) with a minimum one-sided significance of $2.1\sigma$ and a Bayes factor of 10.18~\cite{JUNO:2026tap}. This preference arises from the combined analysis rather than from JUNO data alone.

In Ref.~\cite{Kang:2014mka}, four minimally modified TBM mixing patterns, denoted A--D, were found to be compatible with the oscillation data available at that time.\footnote{Earlier work also considered the bimaximal (BM)-based cases E and F; the present reassessment is restricted to the four TBM-based cases A--D.}
Cases A and B preserve the first and second columns of the TBM mixing matrix and lead to exact solar-reactor mixing angle relations. Cases C and D instead preserve the $\tau$ and $\mu$ rows, respectively, leading to exact atmospheric-reactor relations and predicting opposite octants of $\theta_{23}$. Unlike in cases A and B, the solar mixing angle in cases C and D is not fixed by the reactor mixing angle but enters directly into the corresponding Dirac CP phase sum rules. The improved JUNO measurement can therefore have qualitatively different effects: it strengthens the constraints on the exact mixing relations of cases A and B, while offering the possibility of more precise predictions for $\delta_{\rm CP}$ in cases C and D.

In this work, we revisit these four mixing patterns using the updated JUNO results \cite{JUNO:2026tap} and the NuFIT 6.1 global analysis \cite{NuFIT:2025v61}. We examine the correlations in the $(\sin^2\theta_{12},\delta_{\rm CP})$ plane and quantitatively test the solar-reactor relations of cases A and B and the atmospheric-reactor relations of cases C and D using the corresponding correlated NuFIT profiles for both mass orderings. For cases A and B, we further distinguish the effects of the shift in the solar-angle central value and the reduction in its uncertainty. Since NuFIT 6.1 already incorporates the first JUNO data sample, which overlaps with the updated sample, we treat these inputs separately rather than combining their likelihoods.
Beyond these phenomenological tests, an important question
is whether the corresponding mixing sum rules can be realized
within a concrete neutrino mass model.

In Ref.~\cite{Ahn:2022ufs}, the modular $A_4$ seesaw model
was shown to accommodate case A, while the realization
of cases C and D within the same model was not explored.
Given the stronger JUNO constraint on the exact case A
relation, it is particularly interesting to investigate
whether different choices of the existing model parameters
can realize cases C and D without modifying the mass matrices
or introducing additional parameters.
We therefore re-examine the modular $A_4$ texture of
Ref.~\cite{Ahn:2022ufs} for NO, imposing the complete mixing
conditions of cases A, C, and D, including their Dirac CP
phase sum rules.
We compare the conditional minima and the predicted CP phase
intervals obtained from the exact sum rules for three
solar-angle inputs: NuFIT 6.1, the first JUNO measurement,
and the updated JUNO measurement.
In particular, we identify new parameter solutions realizing
cases C and D and demonstrate that all three mixing patterns
can attain their corresponding conditional minima within
the same modular texture.
We further examine their CP phase solutions using the NuFIT
atmospheric-CP profile and investigate whether the neutrino
mass observables are uniquely determined at fixed oscillation
observables.

The mixing correlations investigated in this work highlight the complementarity of current and future neutrino oscillation experiments. While JUNO provides increasingly precise information on the solar mixing angle, future measurements of $\delta_{\rm CP}$ and $\theta_{23}$ at DUNE and T2HK will offer complementary tests of the modified TBM mixing patterns, particularly cases C and D, which predict opposite atmospheric octants close to maximal mixing.

\section{Minimally modified TBM mixing}
\label{sec:mixing}


In the standard parametrization, the PMNS matrix is written as
\begin{equation}
U^{\rm PMNS} =
\begin{pmatrix}
 c_{12}c_{13} & s_{12}c_{13} & s_{13}e^{-i\delta_{\rm CP}}\\
 -s_{12}c_{23}-c_{12}s_{23}s_{13}e^{i\delta_{\rm CP}}
 &
 c_{12}c_{23}-s_{12}s_{23}s_{13}e^{i\delta_{\rm CP}}
 &
 s_{23}c_{13}\\
 s_{12}s_{23}-c_{12}c_{23}s_{13}e^{i\delta_{\rm CP}}
 &
 -c_{12}s_{23}-s_{12}c_{23}s_{13}e^{i\delta_{\rm CP}}
 &
 c_{23}c_{13}
\end{pmatrix}P,
\label{eq:pmns}
\end{equation}
where $s_{ij}\equiv\sin\theta_{ij}$, $c_{ij}\equiv\cos\theta_{ij}$, and $P$ contains the Majorana phases. These phases do not affect neutrino oscillations and do not enter the Dirac CP phase $\delta_{\rm CP}$.
The leading order mixing matrix retained here is the TBM matrix:
\begin{equation}
U_0^{\rm TBM} =
\begin{pmatrix}
\sqrt{\frac{2}{3}} & \frac{1}{\sqrt{3}} & 0\\
-\frac{1}{\sqrt{6}} & \frac{1}{\sqrt{3}} & \frac{1}{\sqrt{2}}\\
\frac{1}{\sqrt{6}} & -\frac{1}{\sqrt{3}} & \frac{1}{\sqrt{2}}
\end{pmatrix}.
\label{eq:tbm}
\end{equation}
This TBM matrix predicts a vanishing reactor angle $\theta_{13}$ at leading order and therefore requires modification to accommodate the measured nonzero value of $\theta_{13}$ \cite{DoubleChooz:2011ymz,DayaBay:2012fng}.
We consider complex rotations in the $(i,j)$ plane. The four modified mixing matrices (A--D) considered in this work are given by
\begin{align} \label{eq:cases-tbm}
V_{\rm A}&=U_0^{\TBM}U_{23}(\vartheta,\varphi), \qquad 
V_{\rm B}=U_0^{\TBM}U_{13}(\vartheta,\varphi), \nonumber \\ 
V_{\rm C}&=U^\dagger_{12}(\vartheta,\varphi)U_0^{\TBM}, \qquad
V_{\rm D}=U^\dagger_{13}(\vartheta,\varphi)U_0^{\TBM}.
\end{align}
Here $\vartheta$ and $\varphi$ denote the rotation angle and its associated phase.
All four cases represent different modifications of TBM mixing.
Cases A and B preserve the first and second columns of the TBM matrix, respectively, whereas cases C and D preserve its $\tau$ and $\mu$ rows, respectively,
up to unphysical phase redefinitions.
These fixed-column or fixed-row conditions lead to the mixing angle relations and Dirac CP phase sum
rules derived below. 

\subsection{Correlations of solar and atmospheric mixing angles with reactor mixing angle}
\label{subsec:angle-correlation-inputs}

The equivalence of the modified mixing matrices with the standard parametrization in Eq.~(\ref{eq:pmns}) gives relations among the physical mixing angles. For cases A and B, the solar and reactor angles are correlated as
\begin{equation}
\sin^2\theta_{12} =
\begin{cases}
\displaystyle
1-\frac{2}{3(1-\sin^2\theta_{13})},
& \text{Case A},\\[3mm]
\displaystyle
\frac{1}{3(1-\sin^2\theta_{13})},
& \text{Case B}.
\end{cases}
\label{eq:solar-reactor}
\end{equation}
For the numerical analysis, it is useful to invert these relations, which are given by
\begin{equation}
\sin^2\theta_{13} =
\begin{cases}
\displaystyle
1-\frac{2}{3(1-\sin^2\theta_{12})},
& \text{Case A},\\[3mm]
\displaystyle
1-\frac{1}{3\sin^2\theta_{12}},
& \text{Case B}.
\end{cases}
\label{eq:solar-reactor-inverse}
\end{equation}
Thus, the experimentally allowed interval of the reactor mixing angle $\theta_{13}$ translates directly into a restricted interval of the solar mixing angle $\theta_{12}$ for each of these two cases.

For cases C and D, $\theta_{23}$ and $\theta_{13}$ are correlated, and the corresponding relations are
\begin{equation}
\sin^2\theta_{23} =
\begin{cases}
\displaystyle
1-\frac{1}{2(1-\sin^2\theta_{13})},
& \text{Case C},\\[3mm]
\displaystyle
\frac{1}{2(1-\sin^2\theta_{13})},
& \text{Case D}.
\end{cases}
\label{eq:atm-reactor}
\end{equation}
For the measured small value of $\sin^2\theta_{13}$, case C predicts $\sin^2\theta_{23}<1/2$, whereas case D predicts $\sin^2\theta_{23}>1/2$. These two cases therefore provide a direct connection between $\theta_{13}$ and the $\theta_{23}$ octant.


\subsection{Dirac CP phase}

The Dirac CP phase $\delta_{\rm CP}$ can be obtained by comparing the modified mixing matrices with the standard PMNS parametrization. The resulting expressions are conveniently written in terms of $\cos\delta_{\rm CP}$. For case A, we obtain
\begin{equation}
\cos\delta_{\rm CP} =
-\frac{1}{2\tan 2\theta_{23}} \left(
\frac{1-5s_{13}^{\,2}}
{s_{13}\sqrt{2-6s_{13}^{\,2}}}\right).
\label{eq:cosA}
\end{equation}
For case B, the corresponding expression is given by
\begin{equation}
\cos\delta_{\rm CP} =
\frac{2-4s_{13}^{\,2}}
{s_{13}\sqrt{2-3s_{13}^{\,2}}}\,\eta_{23}.
\label{eq:cosB}
\end{equation}
For cases C and D, respectively, we have
\begin{align}
\cos\delta_{\rm CP}\big|_{\rm C}
&=
\frac{s_{13}^{\,2}-(1-3s_{12}^{\,2})(1-3s_{13}^{\,2})}
{3s_{13}\sqrt{1-2s_{13}^{\,2}}\,\Xi},
\label{eq:cosC}\\
\cos\delta_{\rm CP}\big|_{\rm D}
&=
\frac{(1-3s_{13}^{\,2})(1-3s_{12}^{\,2})-s_{13}^{\,2}}
{3s_{13}\sqrt{1-2s_{13}^{\,2}}\,\Xi}.
\label{eq:cosD}
\end{align}
Here we have defined
\begin{equation}
\eta_{ij}\equiv\frac{1}{2\tan2\theta_{ij}},
\qquad
\Xi\equiv\sin2\theta_{12}.
\label{eq:definitions}
\end{equation}
The two solutions for $\delta_{\rm CP}$ are
\begin{equation}
\delta_{\rm CP}=\arccos(\cos\delta_{\rm CP}),
\qquad
\delta_{\rm CP}=2\pi-\arccos(\cos\delta_{\rm CP}),
\label{eq:phase-branches}
\end{equation}
where $0\leq\delta_{\rm CP}<2\pi$. A solution is physical only if
\begin{equation}
|\cos\delta_{\rm CP}|\leq1.
\label{eq:physicality}
\end{equation}
Parameter points that do not satisfy this condition are discarded in our numerical scan.

The relations in Eqs.~(\ref{eq:cosA})--(\ref{eq:cosD}) are independent of the absolute neutrino mass scale and mass ordering. Consequently, the mixing sum rules themselves do not distinguish NO from IO.
However, their compatibility with oscillation data can depend on the ordering through the corresponding experimental constraints.
We therefore consider both orderings in the phenomenological analysis of Sec.~\ref{sec:results}, whereas the numerical study of the modular $A_4$ texture in Sec.~\ref{sec:modularA4} is restricted to NO only.
\section{Numerical Results and Discussion}
\label{sec:results}

We use the updated JUNO measurement as the reference input
for the solar mixing angle $\theta_{12}$ in both mass orderings,
since the JUNO solar oscillation parameters are practically
insensitive to the mass ordering hypothesis~\cite{JUNO:2026tap}.
The quoted central value and $1\sigma$ uncertainty in
Eq.~\eqref{eq:juno-new} give
\begin{equation}
0.2972\leq\sin^2\theta_{12}\leq0.3100
\qquad (1\sigma).
\label{eq:juno-1sigma}
\end{equation}
For the correlation plots, we also show the nominal
$3\sigma$ interval obtained by Gaussian extrapolation
of the quoted $1\sigma$ uncertainty,
\begin{equation}
0.2844\leq\sin^2\theta_{12}\leq0.3228
\qquad (3\sigma).
\label{eq:juno-3sigma}
\end{equation}
This interval is used for graphical illustration and
should not be interpreted as a separately reported
JUNO profile-likelihood interval.
For displaying the $\delta_{\rm CP}$ range, we rely on NuFIT 6.1 data, which is given by
\begin{align}
    \text{NO: } \qquad &  \delta_{\rm CP}=212^{\circ} \ \text{(BF)}, \qquad 176^{\circ}\leq \delta_{\rm CP} \leq 238^{\circ} \ (1\sigma), \qquad 125^{\circ}\leq \delta_{\rm CP} \leq 365^{\circ} \ (3\sigma).
    \nonumber \\
    \text{IO: } \qquad & \delta_{\rm CP}=274^{\circ} \ \text{(BF)}, \qquad 249^{\circ}\leq \delta_{\rm CP} \leq 296^{\circ} \ (1\sigma), \qquad 203^{\circ}\leq \delta_{\rm CP} \leq 335^{\circ} \ (3\sigma).
\end{align}
To consider input of $\theta_{13}$ and $\theta_{23}$, we use the $3\sigma$ ranges from the NuFIT 6.1 data~\cite{Esteban:2026phq}.
The corresponding $3\sigma$ ranges are as follows
\begin{align}
\text{NO: } \qquad \quad &0.02064\leq\sin^2\theta_{13}\leq0.02418,
\quad 
0.435\leq\sin^2\theta_{23}\leq0.584 \quad (3\sigma).  \nonumber \\
\text{IO: }  \qquad \quad &0.02093\leq\sin^2\theta_{13}\leq0.02441,
\quad 
0.439\leq\sin^2\theta_{23}\leq0.584 \quad (3\sigma). 
\label{eq:s23-range}
\end{align}
These NuFIT intervals are considered only for $\theta_{13}$ and $\theta_{23}$. The $\theta_{12}$ constraint is taken directly from the updated JUNO result in Eq.~(\ref{eq:juno-new}), rather than from the global-fit $\theta_{12}$ interval.


For case A and case B, we scan $\sin^2\theta_{12}$ and vary $\theta_{23}$ independently over the interval in Eq.~(\ref{eq:s23-range}). The corresponding value of $\sin^2\theta_{13}$ is obtained from Eq.~(\ref{eq:solar-reactor-inverse}). Points for which the reactor angle falls outside Eq.~(\ref{eq:s23-range}) are discarded. We then evaluate $\cos\delta_{\rm CP}$ using Eqs.~(\ref{eq:cosA}) and (\ref{eq:cosB}), retaining only the points that satisfy the physicality condition in Eq.~(\ref{eq:physicality}), i.e.,  $|\cos\delta_{\rm CP}|\leq1$.
For cases C and D, we scan $\sin^2\theta_{12}$ and $\sin^2\theta_{13}$ over their respective intervals. The atmospheric angle is calculated from Eq.~(\ref{eq:atm-reactor}) and is not varied independently. We retain only the points satisfying the physicality condition $|\cos\delta_{\rm CP}|\leq1$. The two phase branches in Eq.~(\ref{eq:phase-branches}) are shown for each surviving point.
Across the panels for cases A--D, the horizontal axis represents $\sin^2\theta_{12}$ and the vertical axis represents $\delta_{\rm CP}$ in degrees. The color of each point indicates the corresponding value of $\theta_{23}$. 
For cases A and B, $\theta_{23}$ is independently varied, while for cases C and D it is derived from the $\sin^2\theta_{13}$ value through Eq.~(\ref{eq:atm-reactor}).

The scan in $\sin^2\theta_{12}$ is performed over a broader interval than the JUNO allowed region in order to display the full theoretical correlations. The JUNO bands are overlaid to identify the portions of these correlations compatible with the updated measurement. The portions of the correlations outside the displayed
JUNO bands lie beyond the corresponding nominal
solar-angle intervals.



 \FloatBarrier
\subsection{Cases A and B: solar and reactor correlations}

\begin{figure}[!h]
\centering
\includegraphics[width=0.48\textwidth]{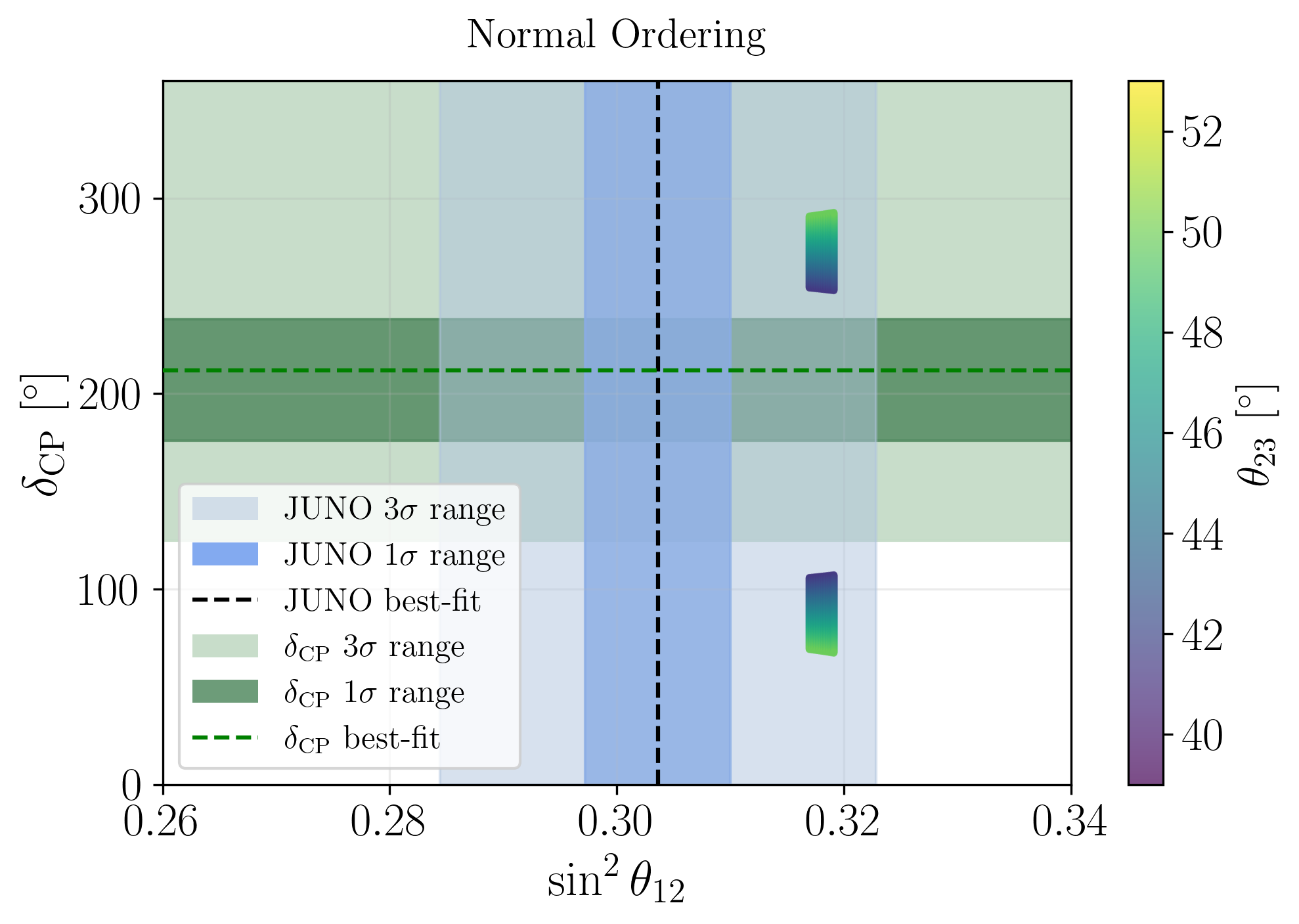}
\hfill
\includegraphics[width=0.48\textwidth]{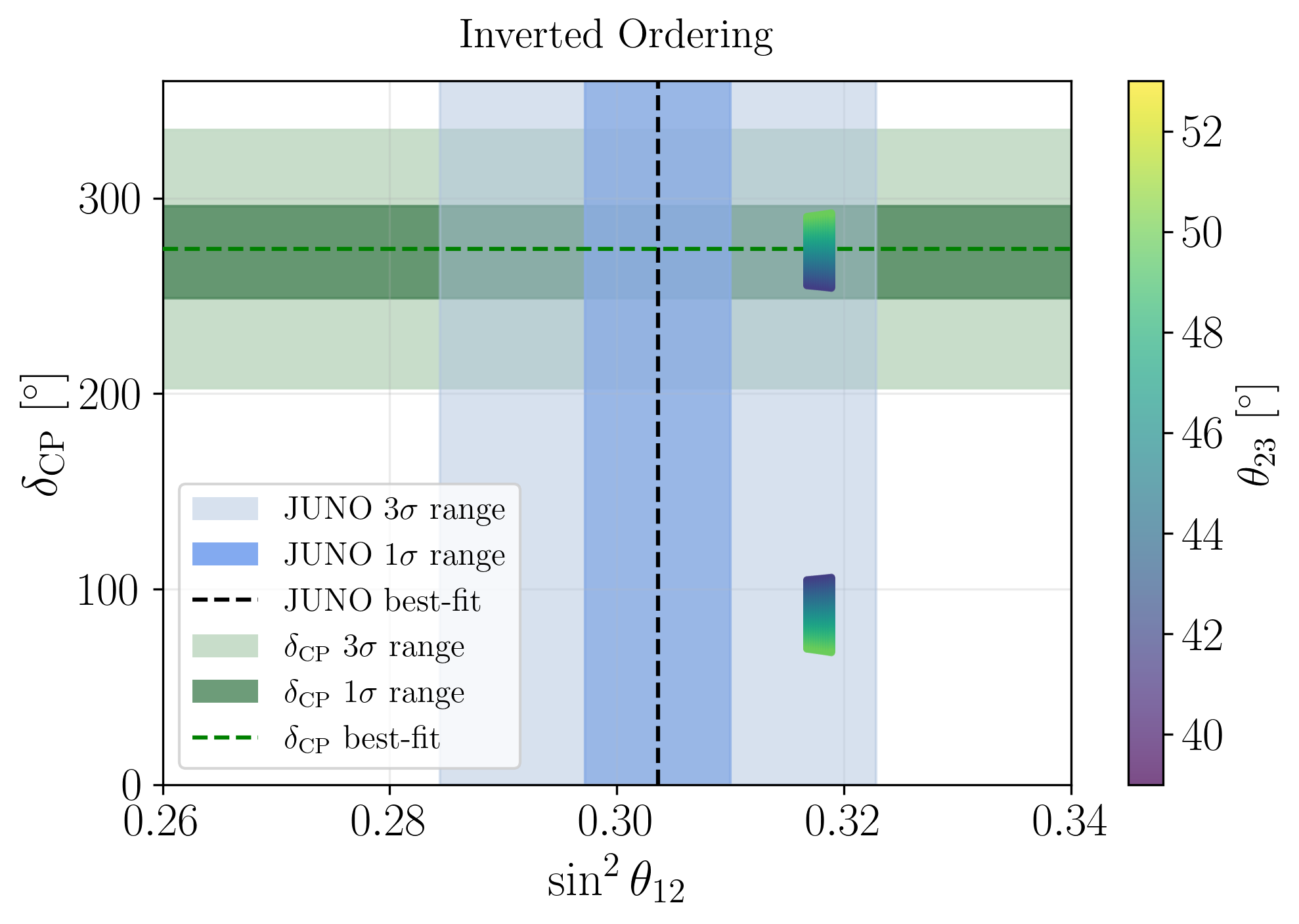}
\caption{Predictions for the Dirac CP phase $\delta_{\rm CP}$ as a function of $\sin^2\theta_{12}$ in case A. The left and right panels correspond to the NO and IO, respectively. The color scale represents $\theta_{23}$ variation. The vertical dashed line denotes the updated JUNO best-fit value \cite{JUNO:2026tap}, while the shaded regions indicate the Gaussian extrapolated $1\sigma$ and $3\sigma$ intervals. The horizontal dashed line and shaded regions correspond to the best-fit value and $1\sigma$, $3\sigma$ regions of $\delta_{\rm CP}$ taken from NuFIT 6.1, respectively \cite{NuFIT:2025v61}.}
\label{fig:A}
\end{figure}

\begin{figure}[!h]
\centering
\includegraphics[width=0.48\textwidth]{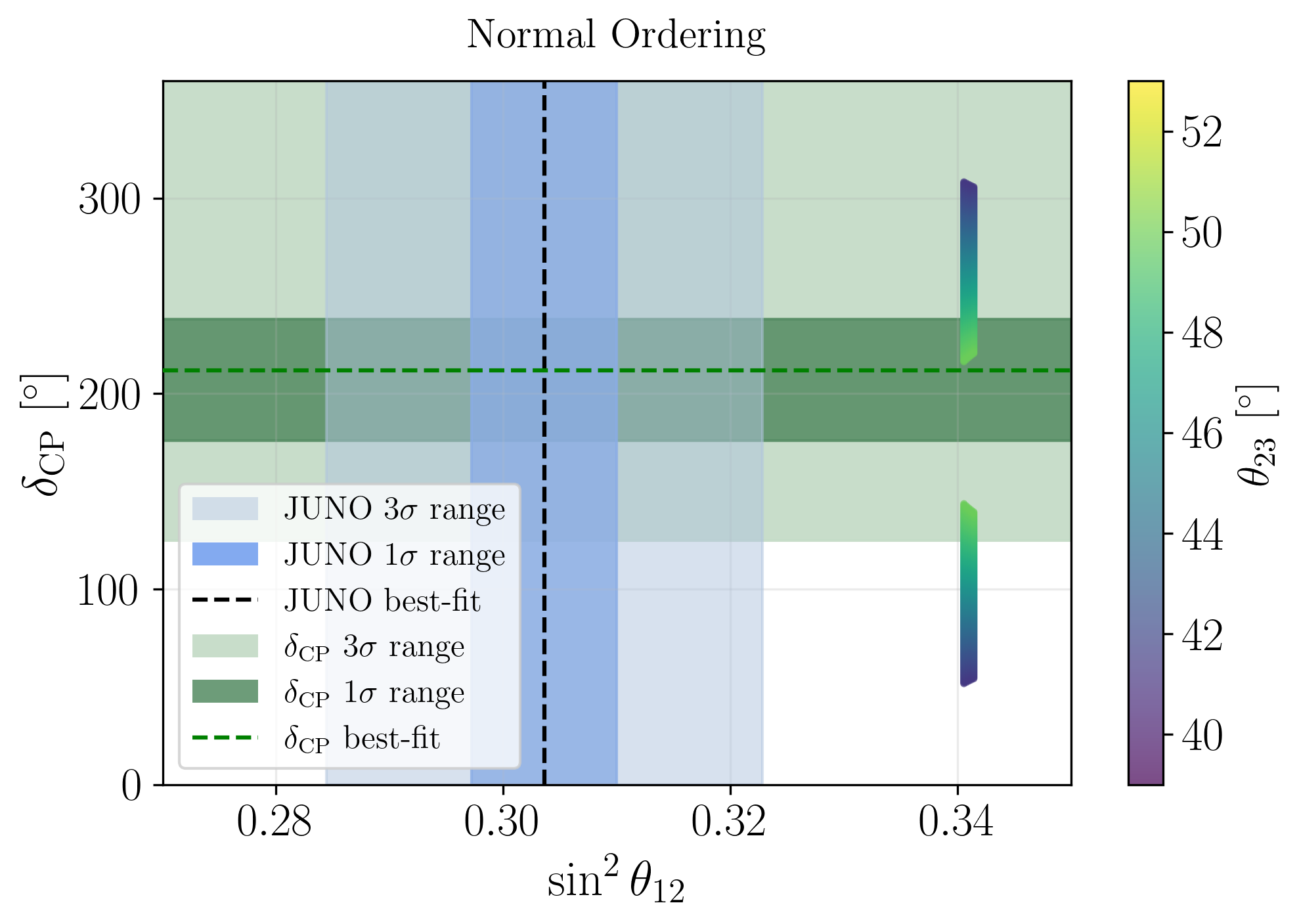}
\hfill
\includegraphics[width=0.48\textwidth]{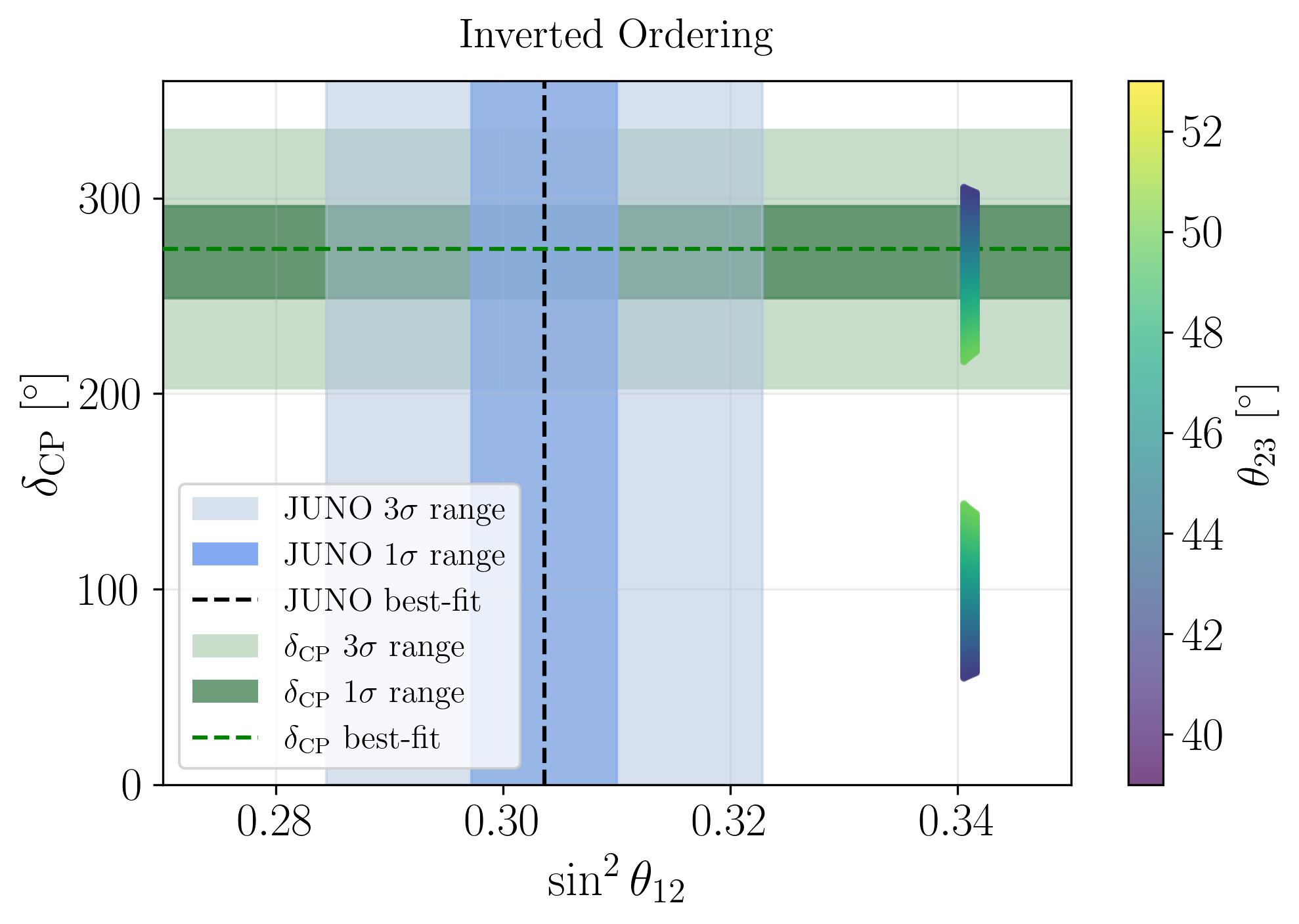}
\caption{Predictions for the Dirac CP phase $\delta_{\rm CP}$ versus $\sin^2\theta_{12}$ in case B for both orderings. The color coding and representation styles are the same as in FIG.~\ref{fig:A}. The case B solar-angle prediction lies outside the JUNO $3\sigma$ interval.}
\label{fig:B}
\end{figure}

Cases A and B are particularly sensitive to the improved JUNO determination because mixing angles $\theta_{12}$ and $\theta_{13}$ are directly correlated as obtained in Eq.~(\ref{eq:solar-reactor}). Combining the $\theta_{13}$ interval in Eq.~(\ref{eq:s23-range}) with these relations gives
\begin{equation}
\begin{aligned}
\mathrm{NO}: \quad&
0.3168\lesssim\sin^2\theta_{12}\lesssim0.3193
\quad\text{(Case A)},\\
&
0.3404\lesssim\sin^2\theta_{12}\lesssim0.3416
\quad\text{(Case B)},\\
\mathrm{IO}: \quad&
0.3167\lesssim\sin^2\theta_{12}\lesssim0.3191
\quad\text{(Case A)},\\
&
0.3405\lesssim\sin^2\theta_{12}\lesssim0.3417
\quad\text{(Case B)}.
\label{eq:caseB-solar}
\end{aligned}
\end{equation}
The comparison with the updated JUNO interval in Eq.~(\ref{eq:juno-3sigma}) leads to a clear distinction between these two cases. 
The predicted Dirac CP phase $\delta_{\rm CP}$ correlations with $\sin^2{\theta_{12}}$ are shown in FIGs.~\ref{fig:A} and \ref{fig:B}. 

The two branches arise from the two solutions of Eq.~(\ref{eq:phase-branches}). The color coding identifies the value of $\theta_{23}$ associated with each point and illustrates how the predicted $\delta_{\rm CP}$ varies with $\theta_{23}$. In case A, the dependence of $\cos\delta_{\rm CP}$ on $\theta_{23}$ allows the phase prediction to be examined over the independently varied $\theta_{23}$ interval. Case B has a similar dependence on $\theta_{23}$ through $\eta_{23}$.
The case A prediction overlaps the Gaussian extrapolated JUNO $3\sigma$ interval, although it lies outside the quoted $1\sigma$ interval. By contrast, the case B prediction lies outside the edge of the JUNO $3\sigma$ interval. Therefore, with the adopted $\theta_{13}$ range, case B is strongly disfavored by the updated $\sin^2{\theta_{12}}$ measurement, 
whereas the case A prediction still overlaps the
Gaussian extrapolated JUNO $3\sigma$ interval.

\subsection{Cases C and D: atmospheric octant correlations}

Unlike cases A and B, cases C and D do not impose a direct $\theta_{12}$--$\theta_{13}$ correlation. The $\theta_{12}$ remains an independent variable in the corresponding expressions for $\cos\delta_{\rm CP}$. Instead, $\theta_{23}$ is correlated with $\theta_{13}$ through Eq.~(\ref{eq:atm-reactor}). This allows the updated JUNO measurements to be used directly to examine the phase predictions in the $(\sin^2\theta_{12},\delta_{\rm CP})$ plane.
The two cases have opposite $\theta_{23}$ octant assignments. Case C predicts $\sin^2\theta_{23}<1/2$, corresponding to the lower octant, whereas case D predicts $\sin^2\theta_{23}>1/2$, corresponding to the upper octant. For the $\theta_{13}$ interval in Eq.~(\ref{eq:s23-range}), the predicted atmospheric angles lie close to maximal mixing. The resulting color variation is therefore expected to be narrower than in cases A and B, where $\theta_{23}$ is scanned independently.
For the reactor-angle ranges in Eq.~\eqref{eq:s23-range}, the predicted values of $\theta_{23}$ lie within approximately
($44.28^\circ$--$44.40^\circ$) for case C and
($45.60^\circ$--$45.72^\circ$) for case D,
including both mass orderings.
\begin{figure}[!h]
\centering
\includegraphics[width=0.48\textwidth]{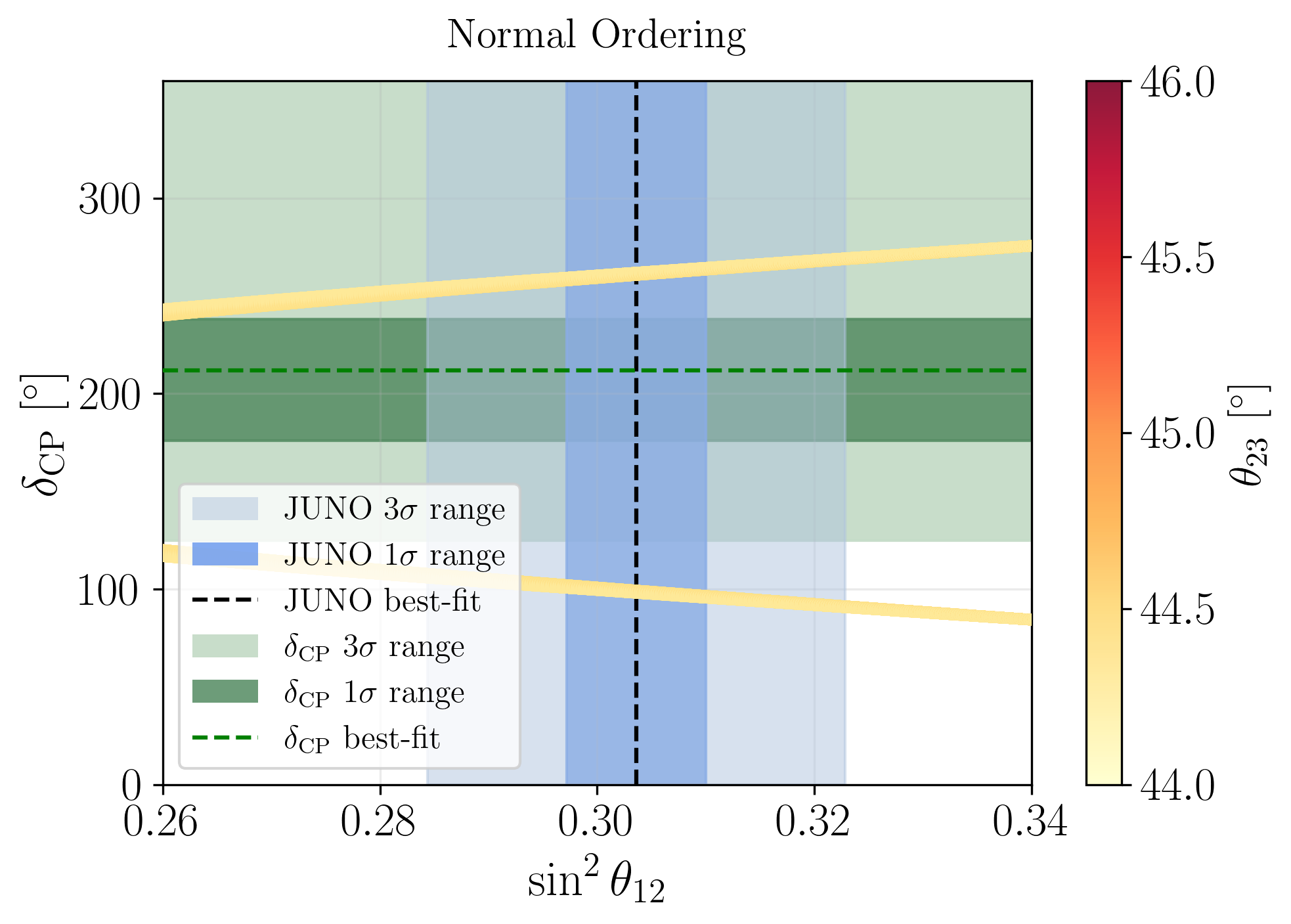}
\hfill
\includegraphics[width=0.48\textwidth]{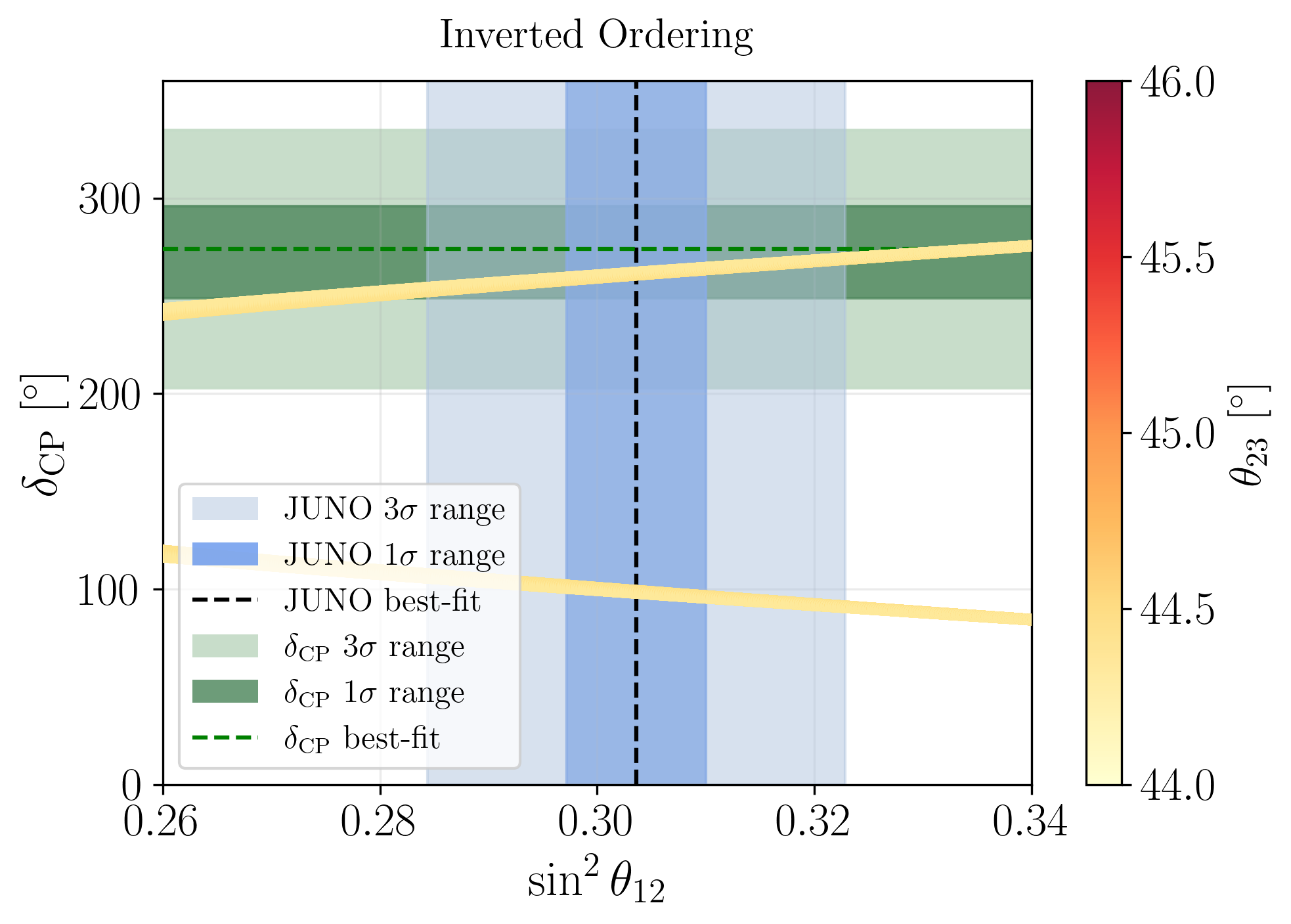}
\caption{Predictions for the Dirac CP phase $\delta_{\rm CP}$ versus $\sin^2\theta_{12}$ for case C and both orderings. The color scale represents $\theta_{23}$; JUNO bands correspond to the quoted uncertainty. Case C predicts the lower atmospheric octant.}
\label{fig:C}
\end{figure}

\begin{figure}[!t]
\centering
\includegraphics[width=0.48\textwidth]{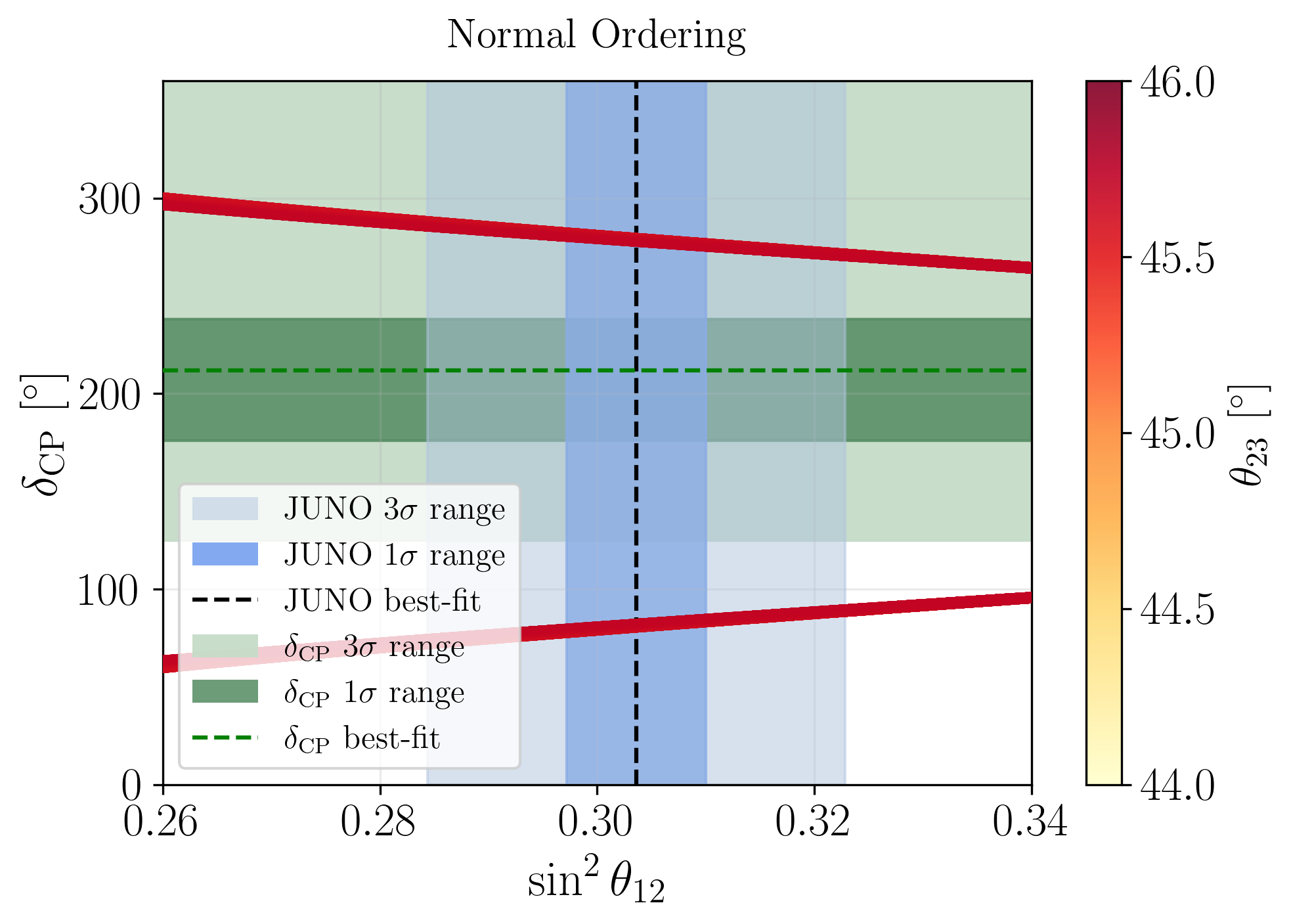}
\hfill
\includegraphics[width=0.48\textwidth]{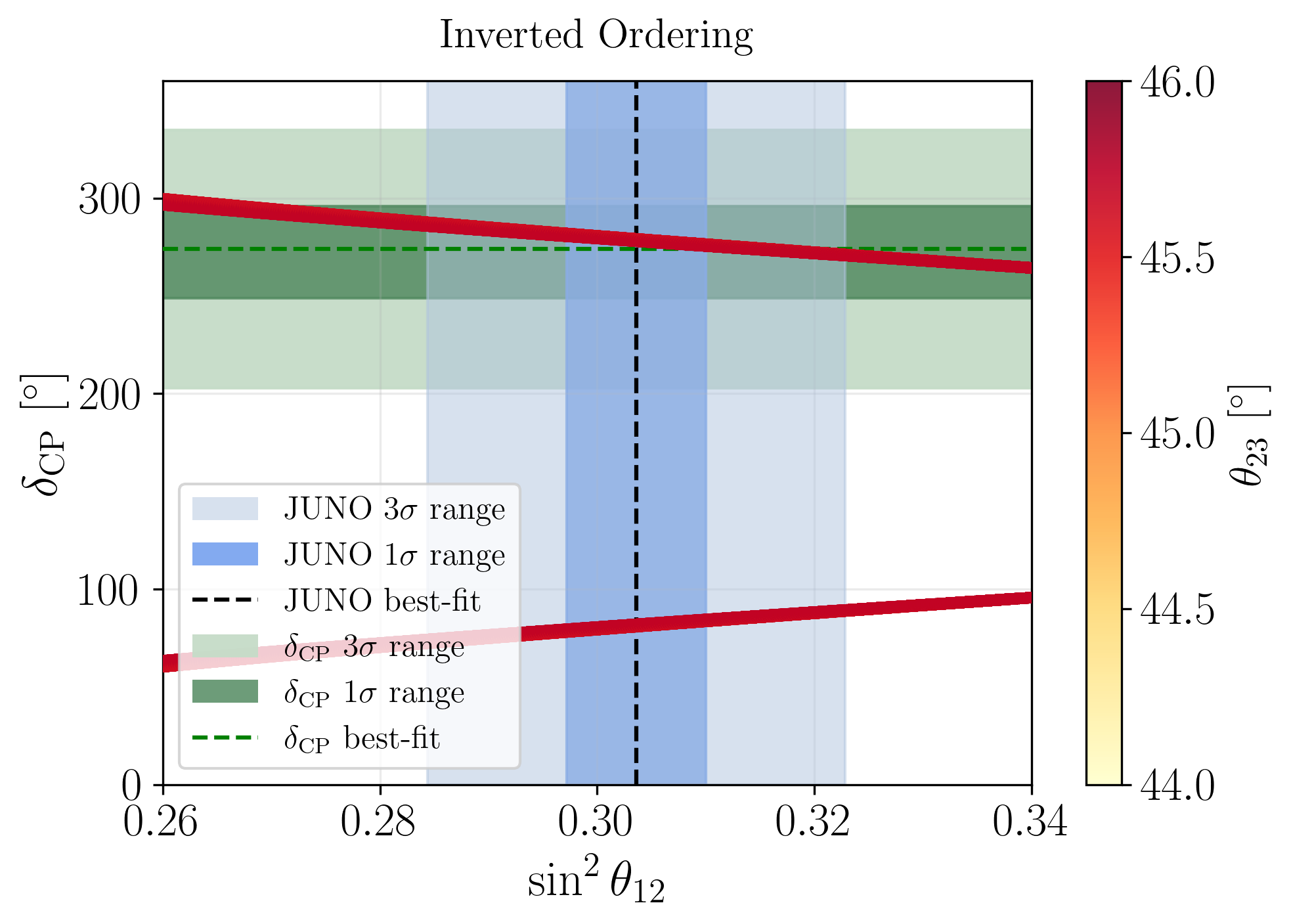}
\caption{Predictions for the Dirac CP phase $\delta_{\rm CP}$ versus $\sin^2\theta_{12}$ for case D and both orderings. The color scale represents $\theta_{23}$; JUNO bands correspond to the quoted uncertainty. Case D predicts the upper atmospheric octant.}
\label{fig:D}
\end{figure}

The $\delta_{\rm CP}$ correlations with $\theta_{12}$ for cases C and D are shown in FIGs.~\ref{fig:C} and \ref{fig:D}. 
Both cases admit physical solutions in the JUNO allowed $\sin^2\theta_{12}$ interval for the adopted $\theta_{13}$ range. Their phase predictions differ because the two expressions in Eqs.~(\ref{eq:cosC}) and (\ref{eq:cosD}) contain opposite signs for the terms involving the $\theta_{12}$ and $\theta_{13}$ angles. The comparison therefore illustrates how the same $\theta_{12}$ measurement can lead to different phase correlations depending on the underlying modified mixing pattern.
These range based comparisons illustrate the implications
of the updated JUNO measurement but do not constitute
a statistical model comparison.

\FloatBarrier

\subsection{Quantitative tests of modified TBM mixing with NuFIT 6.1 and JUNO}
\label{subsec:nufit-comparison}

We now quantitatively test the modified TBM mixing patterns using the correlated NuFIT 6.1 profiles and separately examine the impact of the updated JUNO solar-angle measurement. The NuFIT 6.1 analysis includes Daya Bay's 3158 day spectra and the first 59.1 day JUNO data~\cite{NuFIT:2025v61,DayaBay:2022orm,JUNO:2025gmd}. Because the updated 207.2 day JUNO sample overlaps the earlier one, we do not add the updated JUNO likelihood to the NuFIT profile.

\begin{figure}[!htbp]
\centering
\includegraphics[width=0.88\textwidth]{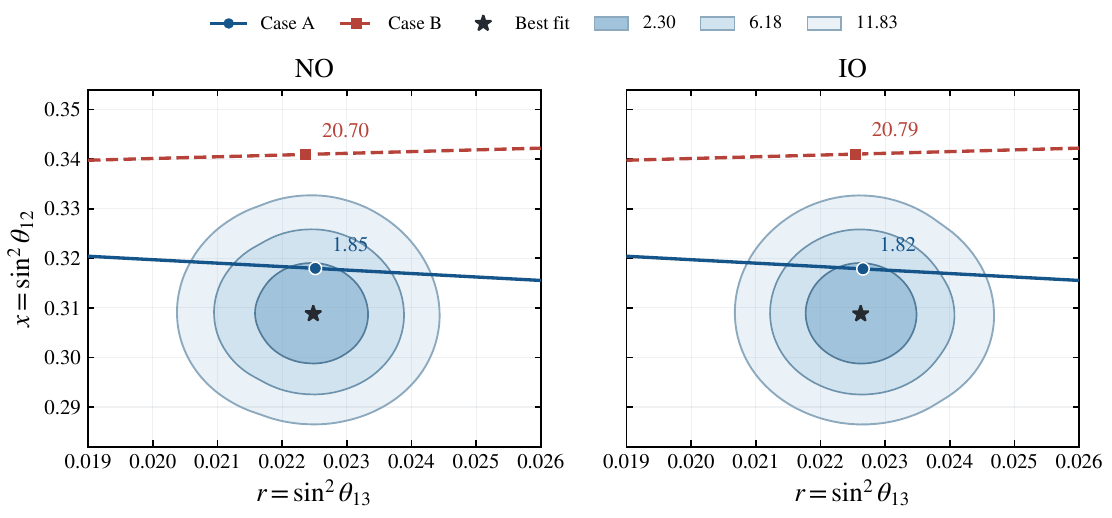}
\caption{NuFIT 6.1 constraints in the
$(r,x)=(\sin^2\theta_{13},\sin^2\theta_{12})$
plane for NO (left) and IO (right).
The contours correspond to $\Delta\chi^2=2.30$, $6.18$,
and $11.83$, with the interpolated profiles normalized
to their continuous minima within each mass ordering.
The solid blue and dashed red curves represent the exact
solar-reactor mixing relations for cases A and B,
respectively.
The circles and squares indicate the minima along
these relations, with their corresponding $\Delta\chi^2$
values, while the stars denote the best-fit points
of the interpolated profiles.}
\label{fig:nufit61-AB-profile}
\end{figure}

\paragraph{\textbf{Tests of cases A and B with NuFIT 6.1 and JUNO:}}
Cases A and B impose exact relations $x=x_k(r)$ ($k=A,B$)
between $x\equiv\sin^2\theta_{12}$ and $r\equiv\sin^2\theta_{13}$,
as given in Eq.~\eqref{eq:solar-reactor}.
We test these relations using the correlated NuFIT 6.1 $(r,x)$
profile, which accounts for correlations between the two mixing
angles while allowing the other oscillation parameters to vary.
For each mass ordering $h$, we evaluate
\begin{equation}
\Delta\chi^2_{k,h}=\min_r\!\left[\Delta\chi^2_h\bigl(r,x_k(r)\bigr)\right]-\min_{r,x}\Delta\chi^2_h(r,x),\qquad k=A,B,
\label{eq:profile-AB-new}
\end{equation}
Each value is measured relative to the best fit within the same ordering, without including the separate NuFIT preference between NO and IO. Table~\ref{tab:profile-AB-new} and FIG.~\ref{fig:nufit61-AB-profile} summarize the results.
\begin{table}[!htbp]
\centering\small
\caption{Profile test of the exact A/B solar-reactor relations on the correlated NuFIT 6.1 $(r,x)$ surface. The reported $\Delta\chi^2_{\rm prof}$ values are within ordering profile differences, not calibrated exclusion significances.}
\label{tab:profile-AB-new}
\begin{tabular}{ccccc}
\hline\hline
Ordering & Case & $x_*$ & $r_*$ & $\Delta\chi^2_{\rm prof}$ \\
\hline
NO & A & 0.31798 & 0.02251 & 1.85 \\
NO & B & 0.34096 & 0.02236 & 20.70 \\
IO & A & 0.31788 & 0.02266 & 1.82 \\
IO & B & 0.34102 & 0.02254 & 20.79 \\
\hline\hline
\end{tabular}
\end{table}
The minimum profile difference is $\Delta\chi^2_{\rm prof}\simeq1.8$ for A and $\simeq20.7$ for B in either ordering. Thus, the exact case B relation is strongly disfavored by this profile, whereas case A remains comparatively compatible. These results do not constitute a formal model-independent exclusion and do not include the updated JUNO measurement.

For the Gaussian comparisons below, we use the
NuFIT 6.1 NO best-fit values and their $1\sigma$
uncertainties, including SK atmospheric data
\cite{NuFIT:2025v61},
\begin{equation}
\begin{aligned}
x &= 0.3088^{+0.0067}_{-0.0066},\\
r &= 0.02248^{+0.00055}_{-0.00059},\\
z &= 0.470^{+0.017}_{-0.014},
\end{aligned}
\label{eq:study_retained_inputs}
\end{equation}
where $z\equiv\sin^2\theta_{23}$.
For the A/B comparison, we use the symmetrized
reactor-angle uncertainty $\sigma_r=0.00057$
and evaluate the first and updated JUNO
solar-angle measurements separately.
For a given solar-angle measurement, we denote
the central value and its $1\sigma$ uncertainty
by $\mu_x$ and $\sigma_x$, respectively, and minimize
\begin{equation}
q_k(\mu_x,\sigma_x)=\min_r\left\{\left(\frac{r-0.02248}{0.00057}\right)^2+\left[\frac{x_k(r)-\mu_x}{\sigma_x}\right]^2\right\},\qquad k=A,B.
\label{eq:juno_matched_gaussian}
\end{equation}
This matched Gaussian comparison is approximate because the reactor-angle input is not necessarily independent of the information entering the JUNO analyses. 
The two JUNO results are not combined with each other or with the NuFIT correlated profile in a joint likelihood analysis.
The quantity $\sqrt{q_k}$ measures the discrepancy between
the exact mixing relation and the experimental inputs
in units of their uncertainties.
For independent Gaussian inputs and a locally linear mixing relation,
it can be approximately interpreted as the number of
standard deviations associated with the discrepancy.
However, since the independence of the inputs is not
guaranteed, it should not be interpreted as a calibrated
exclusion significance.

The updated JUNO result shifts $\mu_x$ from 0.3092 to
0.3036 and reduces $\sigma_x$ from 0.0087 to 0.0064,
corresponding to a 26.4\% reduction in the $1\sigma$
uncertainty.
To distinguish these two effects, we vary the central value
and uncertainty one at a time.
Table~\ref{tab:juno_decomposition} presents the two reported
JUNO inputs and the two hypothetical cases in which either
the solar-angle central value or its uncertainty is changed
separately.
For case A, the increase in $\sqrt{q_A}$ from 1.01 to 2.25
indicates a noticeable increase in tension with the exact
mixing relation.
Case B already exhibits a substantial discrepancy with
the first JUNO result, which becomes considerably larger
with the updated measurement, as reflected by the increase
in $\sqrt{q_B}$ from 3.65 to 5.84.
When the two effects are considered separately, the shift
in the central value has a larger impact on case A,
whereas the improved precision has a larger impact on case B.
\begin{table}[!htbp]
\centering\small
\caption{Gaussian discrepancy $\sqrt{q_k}$ for the exact A/B relations. The intermediate rows independently change the solar-angle central value or uncertainty and do not represent additional measurements. None of the entries is a calibrated exclusion significance.}
\label{tab:juno_decomposition}
\begin{tabular}{cccc}
\hline\hline
Solar-angle input & $(\mu_x,\sigma_x)$ & $\sqrt{q_A}$  & $\sqrt{q_B}$  \\
\hline
First JUNO (59.1 days) & $(0.3092,0.0087)$ & 1.01 & 3.65 \\
Improved precision only & $(0.3092,0.0064)$ & 1.37 & 4.97 \\
Shifted central value only & $(0.3036,0.0087)$ & 1.65 & 4.30 \\
Updated JUNO (207.2 days) & $(0.3036,0.0064)$ & 2.25 & 5.84 \\
\hline\hline
\end{tabular}
\end{table}
These comparisons quantify the solar-angle constraint on the exact A/B relations; their implications for a specific neutrino mass model depend on which mixing patterns it can realize.

\begin{figure}[!htbp]
\centering
\includegraphics[width=\linewidth]{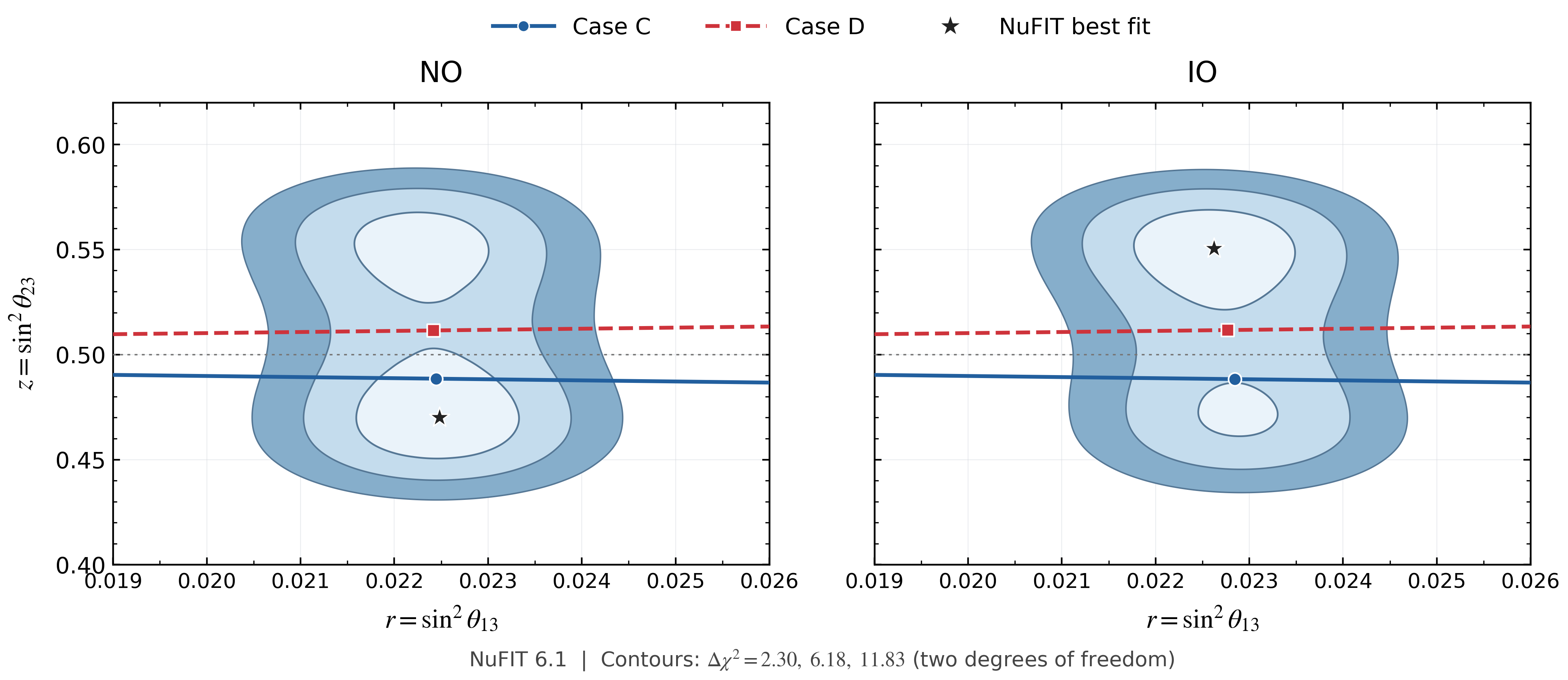}
\caption{NuFIT 6.1 constraints in the $(r,z)=(\sin^2\theta_{13},\sin^2\theta_{23})$ plane for NO (left) and IO (right). 
The shaded regions and contours correspond to
$\Delta\chi^2=2.30$, $6.18$, and $11.83$,
obtained by minimizing over the remaining oscillation
parameters and normalized to the continuous minimum
of the interpolated profile within each ordering.
The blue solid and red dashed curves represent the exact atmospheric-reactor relations for cases C and D, respectively. Circles and squares mark the minimum $\Delta\chi^2$ points along these curves, stars indicate the NuFIT best-fit points, and the dotted horizontal line denotes maximal atmospheric mixing. The additional C/D Dirac CP phase sum rules are not imposed.}
\label{fig:nufit61-CD-profile}
\end{figure}

\paragraph{\textbf{Tests of cases C and D with NuFIT 6.1:}}
For cases C and D, we test the atmospheric-reactor mixing angle relations given in Eq.~(\ref{eq:atm-reactor}) using the correlated NuFIT 6.1 $(r,z)$ profile, where $z\equiv\sin^2\theta_{23}$.
Following the within ordering minimization of Eq.~\eqref{eq:profile-AB-new}, with $x$ replaced by $z$, we find $\Delta\chi^2_{\rm prof}=1.13$ (2.41) for C and 2.61 (2.91) for D in NO (IO). FIG.~\ref{fig:nufit61-CD-profile} shows that both relations intersect the $\Delta\chi^2=6.18$ contour region for either ordering. Unlike the complete C/D sum rules, this test constrains only the atmospheric-reactor relation; $\delta_{\rm CP}$ is freely profiled in the NuFIT surface. We impose the complete C/D conditions in the conditional phase analysis below.
The atmospheric-reactor relations of cases C and D do not
involve the solar mixing angle.
The updated JUNO measurement instead affects their
$\delta_{\rm CP}$ predictions through the additional
CP phase sum rules.
We quantify this effect below using the complete mixing
conditions, independently of a specific neutrino mass model.

\paragraph{\textbf{Impact of JUNO on Dirac CP phase predictions:}}

For this conditional phase analysis, we use the NO inputs
and impose the exact mixing angle and CP phase sum rules
of cases A, C, and D given in Sec.~\ref{sec:mixing}.
Case A is retained to compare its response to the updated
solar-angle measurement with those of cases C and D.
We keep the reactor and atmospheric inputs in
Eq.~\eqref{eq:study_retained_inputs} fixed and
consider three solar-angle inputs separately:
the NuFIT 6.1 value in the same equation,
the first JUNO result in Eq.~\eqref{eq:juno-first},
and the updated result in Eq.~\eqref{eq:juno-new}.

For each solar input $s$, we define the approximate
chi-square function in mixing observable space,
\begin{equation}
Q_s(x,r,z)=
\sum_{O=x,r,z}
\left[\frac{O-\mu_{O,s}}{\sigma_{O,s}}\right]^2.
\label{eq:study_sumrule_objective}
\end{equation}
Here $\mu_{O,s}$ and $\sigma_{O,s}$ are the central
values and uncertainties, with the upper or lower
uncertainty selected according to the sign of the residual.
Only the solar input changes between comparisons.
Experimental information on $\delta_{\rm CP}$ is not
included in $Q_s$.
This factorized approximation neglects correlations
among the adopted inputs, including those associated
with the external reactor information used by JUNO.
As in the A/B comparison above, the overlapping solar
measurements are not combined in a joint likelihood.

\begin{figure*}[t]
\centering
\includegraphics[width=\textwidth]
{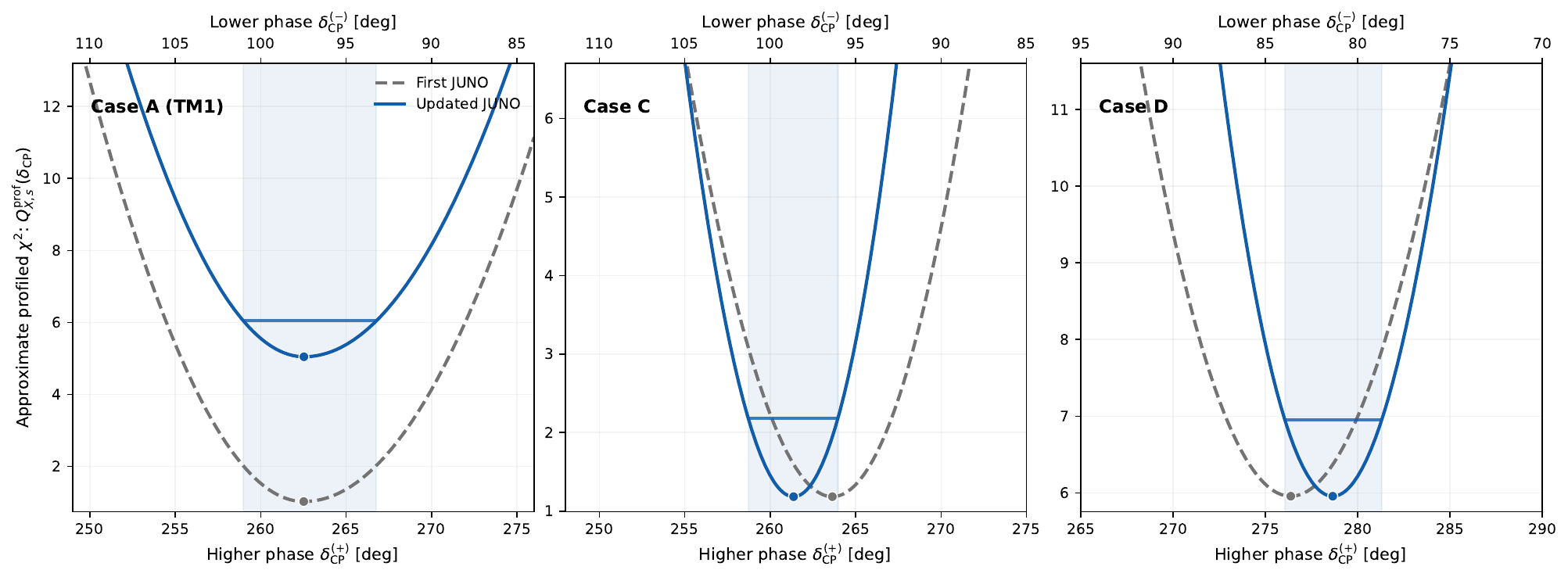}
\caption{Phase-dependent profiles for cases
A, C, and D obtained from the exact mixing
sum rules with the adopted NO inputs, using the first
JUNO (gray dashed) and updated JUNO (blue solid)
solar-angle measurements.
The profiles are defined in mixing observable space
by Eq.~\eqref{eq:study_conditional_profile}.
The bottom and top horizontal axes show the
higher- and lower-phase solutions,
respectively, related by
$\delta_{\rm CP}^{(-)}
=360^\circ-\delta_{\rm CP}^{(+)}$.
Dots indicate the minima, while the blue bands
and horizontal segments denote the intervals
defined by $\Delta Q_{X,s}\leq1$ for the
updated JUNO input.}
\label{fig:study_case_response}
\end{figure*}

Let $\mathcal S_X(\delta)$ be the set of physical
$(x,r,z)$ values satisfying the complete mixing angle
and CP phase sum rules of case $X=A,C,D$ at
$\delta_{\rm CP}=\delta$.
We define
\begin{equation}
\begin{aligned}
Q^{\mathrm{prof}}_{X,s}(\delta)
&=\min_{(x,r,z)\in\mathcal S_X(\delta)}Q_s(x,r,z),\\
Q^{\min}_{X,s}
&=\min_{\delta}Q^{\mathrm{prof}}_{X,s}(\delta),\\
\Delta Q_{X,s}(\delta)
&=Q^{\mathrm{prof}}_{X,s}(\delta)-Q^{\min}_{X,s}.
\end{aligned}
\label{eq:study_conditional_profile}
\end{equation}
The condition $\Delta Q_{X,s}\leq1$ defines the
approximate conditional phase intervals used below.
Neither these intervals nor the minimum $Q_s$ values
are calibrated experimental confidence statements.
They also differ from the correlated NuFIT profile
tests above, which impose only the individual mixing
angle relations and freely profile over $\delta_{\rm CP}$.
Table~\ref{tab:study_matched_minima} presents the
minimum values $Q^{\min}_{X,s}$, the two CP phase
solutions at each minimum, and the width of each
CP phase interval defined by $\Delta Q_{X,s}(\delta)\leq1$.
FIG.~\ref{fig:study_case_response} compares the profiles
for the first and updated JUNO inputs.
Since the sum rules determine $\cos\delta_{\rm CP}$,
each solution has a partner satisfying
$\delta_{\rm CP}^{(-)}=360^\circ-\delta_{\rm CP}^{(+)}$,
where the superscripts denote the lower- and higher-phase
branches, respectively.
The two branches have equal $Q_s$ values and interval
widths because experimental CP phase information is
not included in this comparison.

\begin{table*}[t]
\centering\small
\caption{Conditional minima $Q^{\min}_{X,s}$,
the corresponding CP phase solutions, and
the width $W_\delta$ of each phase interval
obtained from the exact mixing sum rules of cases
$X=A,C,D$ with the adopted NO inputs.
The three solar-angle inputs are taken from
NuFIT 6.1, the first JUNO result (JUNO 59.1 days), and the
updated JUNO result (JUNO 207.2 days).
The phase intervals $W_{\delta}$ are defined by
$\Delta Q_{X,s}\leq1$ and are not calibrated
experimental confidence intervals.}
\label{tab:study_matched_minima}
\label{tab:study_phase_intervals}
\begin{tabular}{llrrrr}
\hline\hline
Case & Solar input & $Q^{\min}_{X,s}$
& $\delta_{\rm CP}^{(-)}$ [$^\circ$]
& $\delta_{\rm CP}^{(+)}$ [$^\circ$]
& $W_{\delta}[^\circ$]\\
\hline
A & NuFIT 6.1
& 1.880 & 97.48 & 262.52 & 7.80\\
A & JUNO 59.1 
& 1.022 & 97.48 & 262.52 & 7.81\\
A & JUNO 207.2 
& 5.046 & 97.47 & 262.53 & 7.80\\
\hline
C & NuFIT 6.1
& 1.184 & 96.52 & 263.48 & 5.37\\
C & JUNO 59.1 
& 1.184 & 96.36 & 263.64 & 7.02\\
C & JUNO 207.2 
& 1.184 & 98.63 & 261.37 & 5.26\\
\hline
D & NuFIT 6.1
& 5.957 & 83.47 & 276.53 & 5.38\\
D & JUNO 59.1 
& 5.957 & 83.63 & 276.37 & 7.03\\
D & JUNO 207.2 
& 5.957 & 81.35 & 278.65 & 5.26\\
\hline\hline
\end{tabular}
\end{table*}

For case A, the solar-reactor relation predicts
$x\simeq0.318$, above the updated JUNO central value.
Its minimum therefore increases from 1.022 to 5.046.
The case A CP phase sum rule, however, depends on the
reactor and atmospheric angles after the solar-reactor
relation is imposed.
Consequently, the higher-phase solution remains near
$262.5^\circ$, and its conditional interval width
stays close to $7.8^\circ$.
The updated measurement thus increases the discrepancy
with the exact case A relation while producing little
change in its CP phase prediction.
For cases C and D, the atmospheric-reactor relations
predict $z\simeq0.4885$ and $0.5115$, respectively,
without fixing $x$.
For all three solar inputs considered here, the choice
$x=\mu_{x,s}$ admits a physical CP phase solution near
the minimum.
The solar contribution to $Q_s$ can therefore vanish,
and the minimum is determined by the common reactor
and atmospheric inputs.
This explains why $Q^{\min}_{C,s}=1.184$ and
$Q^{\min}_{D,s}=5.957$ remain unchanged.
The larger minimum for D mainly reflects its greater
separation from the adopted NO atmospheric best fit,
$z=0.470$.

At fixed $\delta_{\rm CP}$, however, the C/D phase sum
rules constrain $x$ and $r$ together.
The shift of the solar central value therefore moves
the preferred CP phases, while the reduced solar
uncertainty narrows their conditional intervals.
The higher-phase minimum shifts from $263.64^\circ$
to $261.37^\circ$ for C and from $276.37^\circ$
to $278.65^\circ$ for D.
The corresponding interval widths decrease from
$7.02^\circ$ and $7.03^\circ$, respectively, to
$5.26^\circ$ for both cases.
The lower-phase solutions shift in the opposite
directions and have the same interval widths.
Thus, the updated JUNO input shifts each C/D phase
solution by about $2.3^\circ$ and reduces its interval
width by approximately $25\%$, without changing the
minimum discrepancy within this approximation.
This is the principal effect of the improved solar-angle
measurement on the complete C/D mixing conditions.

These conditional phase predictions follow from the
exact mixing sum rules and the adopted oscillation
inputs, without assuming a particular neutrino mass
matrix.
In the following section, we examine whether the
corresponding minima and phase-dependent profiles
can be realized within the modular $A_4$ seesaw
model of Ref.~\cite{Ahn:2022ufs}.

\section{JUNO Constraints on the Modular $A_4$ Seesaw Model}
\label{sec:modularA4}

We now investigate the realization of the sum-rule
results in Sec.~\ref{subsec:nufit-comparison} within
the modular $A_4$ seesaw model of Ref.~\cite{Ahn:2022ufs}.
That work presented a realization of case A.
Here we reassess case A with the updated inputs and
identify parameter solutions for cases C and D,
including their complete CP phase sum rules.
All three patterns can be realized using the existing
model parameters while retaining its mass matrices
and CP symmetry assumption.
We consider NO throughout this section.
Case B is omitted because its exact solar-reactor
relation is already strongly disfavored by the
correlated profile test in Sec.~\ref{subsec:nufit-comparison}.

\subsection{Model and numerical framework}
\label{subsec:A4_mixing}

\paragraph{\textbf{Mass matrices and model parameters:}}

In the modular $A_4$ seesaw model of
Ref.~\cite{Ahn:2022ufs}, the charged-lepton mass matrix
is diagonal, and the neutrino sector is determined
by the level-3 modular forms
$Y(\tau)=(Y_1,Y_2,Y_3)$, which transform as an $A_4$
triplet.
Following the published model, we impose CP symmetry
on the coupling constants, so that CP violation
originates from the complex modulus $\tau$.

We define the ratios
\begin{equation}
\begin{gathered}
a_Y=\frac{Y_2}{Y_1},\qquad
b_Y=\frac{Y_3}{Y_1},\\
\beta=\frac{\beta_2}{\beta_1},\qquad
\beta'=\frac{\beta_3}{\beta_1},\qquad
\gamma=\frac{\gamma_2}{\gamma_1},\qquad
\gamma'=\frac{\gamma_3}{\gamma_1},
\end{gathered}
\label{eq:original_texture_ratios}
\end{equation}
where $\beta_i$ and $\gamma_i$ denote the real
couplings in the Dirac and Majorana sectors,
respectively.
With
$
u_Y=1+2a_Yb_Y,\quad
v_Y=1-a_Yb_Y,\quad
w_Y=b_Y^2-a_Y,\quad
t_Y=b_Y^2+2a_Y$,
the dimensionless Dirac and right-handed Majorana
mass matrices are given by
\begin{equation}
\begin{aligned}
D_0&=
\begin{pmatrix}
1&\beta b_Y&\beta'a_Y\\
b_Y&\beta a_Y&\beta'\\
a_Y&\beta&\beta'b_Y
\end{pmatrix},\\[3pt]
H_0&=
\begin{pmatrix}
u_Y+\frac43\gamma v_Y&
2\gamma b_Y&
-\frac23\gamma w_Y+\gamma't_Y\\
2\gamma b_Y&
\frac43\gamma w_Y+\gamma't_Y&
u_Y-\frac23\gamma v_Y\\
-\frac23\gamma w_Y+\gamma't_Y&
u_Y-\frac23\gamma v_Y&
-4\gamma b_Y
\end{pmatrix}.
\end{aligned}
\label{eq:original_texture_matrices}
\end{equation}
The physical mass matrices are
$m_D=\beta_1Y_1v_uD_0$ and
$M_R=\gamma_1Y_1^2v_\chi H_0$, where
$v_u=\langle H_u\rangle$ and
$v_\chi=\langle\chi\rangle$.
The light neutrino mass matrix is obtained through
the seesaw relation,
\begin{equation}
M_\nu=-m_0D_0^TH_0^{-1}D_0,\qquad
m_0=\left|\frac{\beta_1^2v_u^2}
{\gamma_1v_\chi}\right|.
\label{eq:original_texture_seesaw}
\end{equation}
An irrelevant overall phase has been absorbed into the neutrino fields.
Apart from the overall mass scale $m_0$, the
neutrino mass matrices are specified by six
real parameters,
\begin{equation}
\mathbf p=
\left(
\operatorname{Re}\tau,\,
\operatorname{Im}\tau,\,
\beta,\,
\beta',\,
\gamma,\,
\gamma'
\right).
\label{eq:original_texture_parameters}
\end{equation}
The mixing matrix and non-negative neutrino
masses are obtained from the Takagi
diagonalization of $M_\nu$.
We retain the mass matrices and the real coupling
assumption of Ref.~\cite{Ahn:2022ufs} without
introducing additional parameters.

\paragraph{\textbf{Mixing conditions and realization test:}}

To investigate whether the modular texture can
reproduce the complete mixing relations of cases
A, C, and D, we impose the corresponding
fixed-column or fixed-row conditions on the
PMNS matrix.
For case A, the required condition is
\begin{equation}
\bigl(|U_{e1}|^2,|U_{\mu1}|^2,|U_{\tau1}|^2\bigr)
=
\left(\frac23,\frac16,\frac16\right).
\label{eq:study_A_complete}
\end{equation}
For cases C and D, we impose
\begin{equation}
\begin{gathered}
\bigl(|U_{\alpha1}|^2,|U_{\alpha2}|^2,
|U_{\alpha3}|^2\bigr)
=
\left(\frac16,\frac13,\frac12\right),\qquad
\alpha=\tau\ \text{(C)},\quad
\alpha=\mu\ \text{(D)}.
\end{gathered}
\label{eq:study_CD_complete}
\end{equation}
These conditions imply both the mixing angle
relations and the Dirac CP phase sum rules
derived in Sec.~\ref{sec:mixing}.
They are imposed in the parameter search
and are not automatically satisfied for
arbitrary choices of the model parameters.

We use the same mixing angle inputs as in
Sec.~\ref{subsec:nufit-comparison}, together with
the NuFIT 6.1 NO mass-squared differences
\cite{NuFIT:2025v61},
\begin{equation}
\begin{gathered}
\Delta m^2_{31}=2.511\times10^{-3}\,\mathrm{eV}^2,\qquad
\Delta m^2_{21}=(7.537^{+0.094}_{-0.100})
\times10^{-5}\,\mathrm{eV}^2.
\end{gathered}
\label{eq:study_mass_inputs}
\end{equation}
The overall neutrino mass scale is fixed by
the central value of $\Delta m^2_{31}$.
We define the model objective by extending
Eq.~\eqref{eq:study_sumrule_objective} to include
the mass ratio,
\begin{equation}
\begin{aligned}
Q_s^{A_4}(\mathbf p)
&=Q_s\bigl(x(\mathbf p),r(\mathbf p),z(\mathbf p)\bigr)
+\left[\frac{R(\mathbf p)-\mu_R}{\sigma_R}\right]^2,\\
R&=\frac{\Delta m^2_{21}}{\Delta m^2_{31}}.
\end{aligned}
\label{eq:original_matched_refit}
\end{equation}
The central value $\mu_R$ and its asymmetric
uncertainties $\sigma_R$ follow from
Eq.~\eqref{eq:study_mass_inputs}, with
$\Delta m^2_{31}$ held fixed.
The uncertainty is selected according to the sign
of the residual, as in the mixing only objective.
Only the solar-angle input varies between the
three comparisons.
This objective uses the same approximation and
treatment of overlapping inputs described in
Sec.~\ref{subsec:nufit-comparison}.

For each case $X=A,C,D$, let $\mathcal P_X$ denote
the model parameter points satisfying
Eqs.~\eqref{eq:study_A_complete} or
\eqref{eq:study_CD_complete} within the search
domain below and its CP-conjugate counterpart.
At fixed $\delta_{\rm CP}=\delta$, the model profile
satisfies
\begin{equation}
Q^{A_4,\mathrm{prof}}_{X,s}(\delta)
\equiv
\min_{\substack{\mathbf p\in\mathcal P_X\\
\delta_{\rm CP}(\mathbf p)=\delta}}
Q_s^{A_4}(\mathbf p)
\geq Q^{\mathrm{prof}}_{X,s}(\delta).
\label{eq:study_model_profile}
\end{equation}
The inequality follows because the model points
obey the same mixing sum rules and the additional
mass-ratio contribution is non-negative.
To test whether the lower bound is attained, we
search for model parameters reproducing the
minimizing mixing observables together with
$R=\mu_R$.
Finding such a solution establishes equality at
the examined phase value.
These minima are obtained under the exact mixing
conditions of cases A, C, and D, rather than from
an unconstrained fit of the modular model.
To obtain the corresponding model solutions,
we search within the parameter ranges
\begin{equation}
\begin{gathered}
0.05\leq\operatorname{Re}\tau\leq0.49,\qquad
0.75\leq\operatorname{Im}\tau\leq1.50,\\
0.05\leq|\beta|,|\beta'|\leq4,\qquad
-2\leq\gamma,\gamma'\leq2.
\end{gathered}
\label{eq:original_refit_domain}
\end{equation}
The signs of the real Dirac coupling ratios
$\beta$ and $\beta'$ can be absorbed into
lepton field rephasings.
The CP-conjugate solutions are obtained as
described in Sec.~\ref{subsec:A4_CP}.
\subsection{Modular realization of mixing sum rules}
\label{subsec:A4_realization}

For all three solar-angle inputs, we find model
parameter solutions that attain the conditional
minima of cases A, C, and D in
Table~\ref{tab:study_matched_minima}.
For the updated JUNO input, three representative
solutions, BP-A, BP-C, and BP-D, are listed in
Tables~\ref{tab:study_BP_parameters}
and~\ref{tab:study_BP_observables}.
For each benchmark $\mathbf p_X$, we evaluate the
modular forms at its specified $\tau$, construct
$D_0$ and $H_0$ using
Eq.~\eqref{eq:original_texture_matrices}, and
diagonalize the resulting light neutrino mass matrix
in Eq.~\eqref{eq:original_texture_seesaw},
\begin{equation}
U_X^T M_\nu(\mathbf p_X)U_X
=\operatorname{diag}(m_1,m_2,m_3).
\label{eq:study_BP_diagonalization}
\end{equation}
Since the charged-lepton mass matrix is diagonal,
$U_X$ is the PMNS matrix up to unphysical phases.
The mixing angles and Dirac CP phase in
Table~\ref{tab:study_BP_observables} are extracted
from this matrix; the phase is obtained from its
rephasing-invariant combinations.

\begin{table}[!ht]
\centering\small
\caption{Benchmark parameters of the modular
$A_4$ model of Ref.~\cite{Ahn:2022ufs}
satisfying the exact mixing conditions of
cases A, C, and D and attaining the
corresponding conditional minima for
the updated JUNO input.
The displayed parameters are rounded; the oscillation
observables are evaluated using the unrounded solutions.}
\label{tab:study_BP_parameters}
\begin{tabular}{lrrr}
\hline\hline
Parameter & BP-A & BP-C & BP-D\\
\hline
$\operatorname{Re}\tau$ & 0.29010 & 0.29503 & 0.29970\\
$\operatorname{Im}\tau$ & 0.99464 & 0.99626 & 1.02190\\
$\beta$  & -1.70075 & -1.67906 & -1.75341\\
$\beta'$ & 2.10463 & 2.00940 & 2.11003\\
$\gamma$ & 0.29887 & 0.31542 & 0.35938\\
$\gamma'$ & 0.20522 & 0.19263 & 0.13701\\
\hline\hline
\end{tabular}
\end{table}
\begin{table}[!ht]
\centering\small
\caption{Oscillation observables predicted at BP-A,
BP-C, and BP-D for the updated JUNO solar-angle input.
These benchmarks attain the corresponding conditional
minima in Table~\ref{tab:study_matched_minima}.}
\label{tab:study_BP_observables}
\begin{tabular}{lrrr}
\hline\hline
Observable & BP-A & BP-C & BP-D\\
\hline
$x=\sin^2\theta_{12}$ & 0.31795 & 0.30360 & 0.30360\\
$r=\sin^2\theta_{13}$ & 0.02255 & 0.02249 & 0.02245\\
$z=\sin^2\theta_{23}$ & 0.47000 & 0.48850 & 0.51149\\
$\delta_{\rm CP}$ [$^\circ$] & 262.53 & 261.37 & 278.65\\
$R=\Delta m^2_{21}/\Delta m^2_{31}$
& 0.030016 & 0.030016 & 0.030016\\
\hline\hline
\end{tabular}
\end{table}

The resulting PMNS matrices give
\begin{equation}
\begin{aligned}
\bigl(|U_{e1}|^2,|U_{\mu1}|^2,|U_{\tau1}|^2\bigr)_{\rm BP-A}
&\simeq(0.666667,0.166667,0.166667),\\
\bigl(|U_{\tau1}|^2,|U_{\tau2}|^2,|U_{\tau3}|^2\bigr)_{\rm BP-C}
&\simeq(0.166667,0.333333,0.500000),\\
\bigl(|U_{\mu1}|^2,|U_{\mu2}|^2,|U_{\mu3}|^2\bigr)_{\rm BP-D}
&\simeq(0.166667,0.333333,0.500000).
\end{aligned}
\label{eq:study_BP_mixing_check}
\end{equation}
Using the higher-precision benchmark parameters,
the maximum absolute deviations of these squared
moduli from the exact fractions in
Eqs.~\eqref{eq:study_A_complete} and
\eqref{eq:study_CD_complete} are below $10^{-9}$.
These direct checks establish the realization of
the complete A/C/D mixing conditions, including
their Dirac CP phase sum rules, at the respective
model parameter points.
In particular, the C/D row conditions and opposite
atmospheric octants arise from the neutrino
diagonalization in the charged-lepton mass basis,
with the mass matrices and real coupling assumption
retained.
The same benchmarks reproduce $R=\mu_R$, with a
negligible mass-ratio contribution to
$Q_s^{A_4}$.
Evaluating Eq.~\eqref{eq:original_matched_refit}
at these benchmarks reproduces the corresponding
sum-rule minima in Table~\ref{tab:study_matched_minima}
at the quoted precision.
Together with the lower bound in
Eq.~\eqref{eq:study_model_profile}, these explicit
solutions demonstrate that the modular model
attains those conditional minima within numerical
precision.

We also test the phase-dependent profiles of
FIG.~\ref{fig:study_case_response}.
For the first and updated JUNO inputs, model
solutions are found at $0.5^\circ$ intervals over
($250^\circ$--$275^\circ$) for cases A and C and
($265^\circ$--$290^\circ$) for case D.
The endpoints of the conditional intervals,
$\Delta Q_{X,s}=1$, are checked explicitly.
For the NuFIT 6.1 solar input, analogous checks
are performed within its conditional intervals.
At every examined point, the model reproduces
the minimizing mixing observables and mass-ratio
target, attaining the sum-rule lower bound in
Eq.~\eqref{eq:study_model_profile}.
These checks establish numerical realization
at the sampled phases and interval endpoints.
Replacing the solar mass-splitting input with the
updated JUNO value in Eq.~\eqref{eq:juno-new}
does not change the mixing only profiles of
Sec.~\ref{subsec:nufit-comparison}.
We also find model solutions reproducing the
updated mass-ratio target at the conditional
minima and phase interval endpoints.
The increased tension with case A does not by
itself imply a corresponding tension with the
unrestricted model, which also admits cases C and D.

\subsection{CP-Conjugate solutions and consistency with NuFIT}
\label{subsec:A4_CP}

All coupling ratios
$\beta,\beta',\gamma,\gamma'$ are real.
The transformation $\tau\rightarrow-\tau^*$
therefore complex conjugates the modular
forms and the light neutrino mass matrix,
\begin{equation}
Y_i(-\tau^*)=Y_i(\tau)^*,\qquad
M_\nu(-\tau^*)=M_\nu(\tau)^*.
\label{eq:study_CP_conjugation}
\end{equation}
Consequently, the neutrino masses and
mixing angles remain unchanged, whereas
the Dirac CP phase transforms as
$\delta_{\rm CP}\rightarrow
360^\circ-\delta_{\rm CP}$.
The corresponding lower-phase solutions
are thus obtained by CP conjugation of
the higher-phase solutions, extending
the positive $\operatorname{Re}\tau$
domain in Eq.~\eqref{eq:original_refit_domain}
to its CP-conjugate counterpart.
Although the two phase solutions give
identical values of $Q_s^{A_4}$, their compatibility
with the experimental CP phase information
can be different.
As a complementary check, we evaluate
the released NuFIT 6.1 NO three-dimensional
$\Delta\chi^2$ profile in
$(z,\Delta m^2_{31},\delta_{\rm CP})$
at the representative solutions obtained
using the updated JUNO solar-angle input.
This profile is minimized over the other
oscillation parameters and retains the
correlations between the atmospheric
mixing angle, the atmospheric mass-squared
difference, and the Dirac CP phase.
It is not an independent likelihood that
can be combined with $Q_s^{A_4}$.

\begin{figure*}[t]
\centering
\includegraphics[width=\textwidth]
{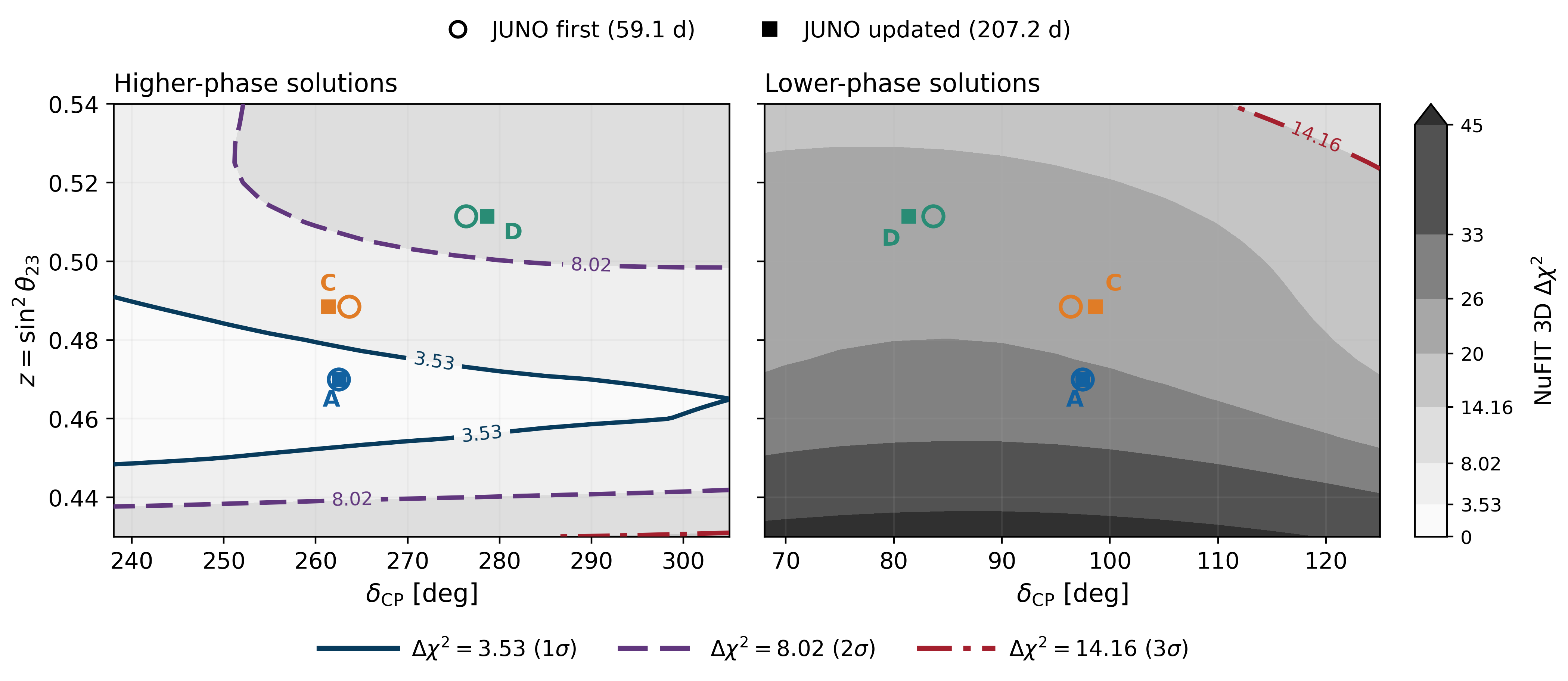}
\caption{NuFIT 6.1 NO three-dimensional atmospheric-CP
profile (IC24 with SK atmospheric data) on the fixed
$\Delta m^2_{31}=2.511\times10^{-3}\,\mathrm{eV}^2$ slice.
The higher- and lower-phase solutions of cases A, C, and D
are shown in the left and right panels, respectively.
Open circles and filled squares correspond to the first
and updated JUNO solar-angle inputs, respectively.
Darker shading indicates larger $\Delta\chi^2$ values.
The contours at $\Delta\chi^2=3.53$, $8.02$, and $14.16$
correspond to the nominal $1\sigma$, $2\sigma$, and $3\sigma$
reference levels for three parameters, rather than
two-dimensional confidence regions on this fixed slice.
The updated JUNO measurement is not included as an
additional likelihood.
}
\label{fig:study_nufit3d_contours}
\end{figure*}

FIG.~\ref{fig:study_nufit3d_contours}
displays the higher- and lower-phase solutions
for the first and updated JUNO inputs on the
fixed $\Delta m^2_{31}$ slice of the
released NuFIT profile.
For the updated JUNO input, the higher-phase solutions
give $\Delta\chi^2=2.66$, $5.07$, and $9.40$ for cases
A, C, and D, respectively, compared with $26.89$,
$24.09$, and $23.00$ for their CP-conjugate counterparts.
The higher-phase solutions lie below the nominal
three-parameter $3\sigma$ reference value of 14.16,
whereas their CP-conjugate counterparts lie above it.
These results indicate better agreement of the
higher-phase solutions with the NuFIT profile
at the examined points, but do not constitute
a calibrated exclusion of the lower-phase branch
over the entire constrained parameter space.

\subsection{Neutrino mass observables}
\label{subsec:A4_masses}

In addition to the mixing angles and the
Dirac CP phase, the modular $A_4$ texture
determines the neutrino masses and Majorana
phases at each parameter point.
We consider three mass observables,
\begin{equation}
\begin{gathered}
\Sigma=\sum_i m_i,\qquad
m_\beta=
\left(\sum_i|U_{ei}|^2m_i^2\right)^{1/2},\\
m_{\beta\beta}
=\left|\sum_iU_{ei}^2m_i\right|
=|(M_\nu)_{ee}|.
\end{gathered}
\label{eq:refit_mass_observables}
\end{equation}
Here $\Sigma$ is the sum of the neutrino
masses, $m_\beta$ is the effective neutrino
mass in beta decay, and $m_{\beta\beta}$
is the effective Majorana mass relevant
to neutrinoless double beta decay
($0\nu\beta\beta$).
None of these quantities is included
in the oscillation fit.

For the benchmark solutions in
Table~\ref{tab:study_BP_parameters},
we obtain
\begin{equation}
\begin{aligned}
(\Sigma,m_\beta,m_{\beta\beta})_A
&=(70.49,11.98,3.59)\;\mathrm{meV},\\
(\Sigma,m_\beta,m_{\beta\beta})_C
&=(69.94,11.73,3.52)\;\mathrm{meV},\\
(\Sigma,m_\beta,m_{\beta\beta})_D
&=(65.86,10.34,2.63)\;\mathrm{meV}.
\end{aligned}
\label{eq:BP_mass_predictions}
\end{equation}
The three benchmark points give neutrino
mass sums close to $0.07\,\mathrm{eV}$,
with effective Majorana masses of a few meV.
These values are specific to the selected
parameter points and do not establish
allowed mass ranges for the corresponding
mixing patterns.

The modular texture also permits different
mass spectra at fixed oscillation observables
when no exact A/C/D sum rule is imposed.
For example, using the updated JUNO solar-angle
central value, the NuFIT 6.1 NO central values
for the remaining mixing angles and mass-squared
differences, and $\delta_{\rm CP}=212^\circ$,
we obtain two solutions with
$(\Sigma,m_{\beta\beta})
=(73.71,1.94)\;\mathrm{meV}$
and $
(\Sigma,m_{\beta\beta})
=(106.86,8.97)\;\mathrm{meV}$.
Thus, even a precise determination of
all neutrino oscillation parameters,
including $\delta_{\rm CP}$, does not
uniquely determine the absolute neutrino
masses or the effective Majorana mass
within the modular $A_4$ texture.

For the standard light Majorana neutrino
exchange mechanism, the neutrinoless
double beta decay rate is given by
\begin{equation}
\left[T^{0\nu}_{1/2}\right]^{-1}
=
G_{0\nu}\,g_A^4\,|\mathcal M_{0\nu}|^2
\left|\frac{m_{\beta\beta}}{m_e}\right|^2,
\label{eq:0nubb-rate-new}
\end{equation}
where $G_{0\nu}$ is the phase-space factor,
$g_A$ is the axial coupling, and
$\mathcal M_{0\nu}$ is the nuclear
matrix element.
The KamLAND-Zen result
$T^{0\nu}_{1/2}>3.8\times10^{26}\,
\mathrm{yr}$ implies
$m_{\beta\beta}<(28$--$122)\,\mathrm{meV}$,
depending on the nuclear matrix
element~\cite{KamLANDZen:2025}.
The projected 10 year nEXO sensitivity
corresponds to approximately
$m_{\beta\beta}=(4.7$--$20.3)\,\mathrm{meV}$
under the same decay
mechanism~\cite{nEXO:2022}.
All three benchmark values in
Eq.~\eqref{eq:BP_mass_predictions}
lie below the current KamLAND-Zen bound
and the most optimistic quoted nEXO
sensitivity.
Nevertheless, these comparisons do not
establish the experimental reach for
the entire parameter space of the
modular texture, since other parameter
solutions can give different neutrino
masses and Majorana phases.



\FloatBarrier

\section{Conclusions}
\label{sec:conclusion}

In this work, we have investigated how the improved determination
of the solar mixing angle by JUNO affects the viability of four
minimally modified TBM mixing patterns, A--D, and their predictions
for the Dirac CP phase.
Using the updated JUNO measurement based on 207.2 live days of data
and the NuFIT 6.1 global analysis, we have quantitatively tested
the exact mixing angle relations of these patterns.
For the exact solar-reactor relations, the correlated NuFIT profiles give $\Delta\chi^2=1.85$ (1.82) for case A and 20.70 (20.79) for case B in NO (IO). Thus, case B is strongly disfavored, while case A remains comparatively compatible. A separate Gaussian comparison shows that the updated JUNO measurement increases the tension for both patterns. For the atmospheric-reactor relations, we obtain $\Delta\chi^2=1.13$ (2.41) for case C and 2.61 (2.91) for case D in NO (IO), indicating that both relations remain compatible with the corresponding NuFIT profiles. These comparisons do not constitute calibrated exclusion tests.

The updated JUNO measurement has qualitatively different implications for the CP phase predictions of cases A, C, and D. For case A, the minimum of our approximate chi-square function increases from 1.022 to 5.046, while the predicted CP phase interval remains nearly unchanged. In contrast, for cases C and D, the corresponding minima remain unchanged at 1.184 and 5.957, respectively, whereas the predicted CP phase solutions shift by approximately $2.3^\circ$ and the widths of their conditional phase intervals decrease by about 25\%. These results reflect the different dependence of the CP phase sum rules on the solar mixing angle.

We have identified parameter solutions within the modular
$A_4$ seesaw model of Ref.~\cite{Ahn:2022ufs} that realize
the complete mixing conditions of cases A, C, and D
and attain their corresponding conditional minima.
In particular, the realization of cases C and D, not
explored within this model in the previous work, requires
neither modifications of the mass matrices nor additional
model parameters.
The increased tension with case A therefore does not
necessarily imply a corresponding tension with the
modular $A_4$ model itself.
A separate comparison with the NuFIT 6.1 atmospheric-CP
profile shows better agreement for the higher-phase
solutions at the examined points.

The effective Majorana masses at BP-A, BP-C, and BP-D
are 3.588, 3.521, and 2.628 meV, respectively, all below
the current KamLAND-Zen bound and the most optimistic
quoted nEXO sensitivity.
These values, however, are specific to the selected
benchmark points. More generally, the modular $A_4$
texture admits substantially different neutrino masses
and effective Majorana masses even at fixed oscillation
observables, demonstrating the limitations of its
predictive power for absolute neutrino masses and
neutrinoless double beta decay.

Our results highlight the complementary roles of JUNO, DUNE, and T2HK in testing lepton mixing sum rules. While JUNO provides increasingly precise constraints on the solar mixing angle, future measurements of $\delta_{\rm CP}$ and $\theta_{23}$ will further test the modified TBM mixing patterns. In particular, distinguishing between cases C and D requires sensitivity to their characteristic CP phase correlations and opposite atmospheric octants close to maximal mixing.


\section*{Acknowledgments} 
\noindent
This work is supported by the National Research Foundation of Korea under Grant NRF-2023R1A2C100609111.


\bibliographystyle{utphys}
\bibliography{references-sk} 
\end{document}